\documentclass[twocolumn, times]{aastex63} 
\usepackage{amsmath}
\usepackage[utf8]{inputenc}

\newcommand{\kms}{\ifmmode {\rm km\ s}^{-1} \else km s$^{-1}$\ \fi}
\newcommand{\ergs}{\ifmmode {\rm erg\ s}^{-1} \else erg s$^{-1}$\ \fi}
\newcommand{\lb}{\ifmmode L_{\rm Bol} \else $L_{\rm Bol}$\ \fi}
\newcommand{\ledd}{\ifmmode L_{\rm Edd} \else $L_{\rm Edd}$\ \fi}
\newcommand{\lx}{\ifmmode L_{\rm 2-10keV} \else  $L_{\rm 2-10keV}$\ \fi}
\newcommand{\ha}{\hbox{H$\alpha$}}
\newcommand{\hb}{\hbox{H$\beta$}}

\newcommand{\mbh}{\ifmmode M_{\rm BH}  \else $M_{\rm BH}$\ \fi}
\newcommand{\lv}{\ifmmode \lambda L_{\lambda}(5100\Ang) \else $\lambda L_{\lambda}(5100\Ang)$\ \fi}
\newcommand{\lbol}{\ifmmode L_{\rm Bol} \else $L_{\rm Bol}$\ \fi}

\newcommand{\nii}{\hbox{[N\,{\sc ii}]}}
\newcommand{\sii}{\hbox{[S\,{\sc ii}]}}
\newcommand{\oiii}{\hbox{[O\,{\sc iii}]}}

\newcommand{\msun}{ M_{\odot}}
\newcommand{\hi}{\hbox{H\,{\sc i}}}

\newcommand{\oh}{\ifmmode 12+ \log({\rm O/H}) \else 12+log(O/H) \fi}
\newcommand{\mdot}{\ifmmode \dot{m} \else \dot{m} \fi }
\newcommand{\llog}{\ifmmode {\rm log} \else {\rm log} \fi }
\newcommand{\Ang}{\mathring{\mathrm{A}}}

\accepted{ by ApJ, April 10, 2021}

\shorttitle{SF and kinematics in NGC 1365}
\shortauthors{Gao et al.}

\begin{document}
\title{The Nuclear Region of NGC1365: Star Formation, Negative Feedback, and Outflow Structure}

\correspondingauthor{Fumi Egusa, Yulong Gao, Guilin Liu}
\email{fegusa@ioa.s.u-tokyo.ac.jp, ylgao@mail.ustc.edu.cn, glliu@ustc.edu.cn}

\author[0000-0002-5973-694X]{Yulong Gao}
\affiliation{CAS Key Laboratory for Research in Galaxies and Cosmology, Department of Astronomy, University of Science and Technology of China, Hefei 230026, China}
\affiliation{School of Astronomy and Space Science, University of Science and Technology of China, Hefei 230026, China}
\affiliation{Institute of Astronomy, The University of Tokyo, Osawa 2-21-1, Mitaka, Tokyo 181-0015, Japan}
\affiliation{Department of Astronomy, Nanjing University, Nanjing 210093, China}
\affiliation{Key Laboratory of Modern Astronomy and Astrophysics (Nanjing University), Ministry of Education, Nanjing 210093, China}

\author[0000-0002-1639-1515]{Fumi Egusa}
\affiliation{Institute of Astronomy, The University of Tokyo, Osawa 2-21-1, Mitaka, Tokyo 181-0015, Japan}

\author[0000-0003-2390-7927]{Guilin Liu}
\affiliation{CAS Key Laboratory for Research in Galaxies and Cosmology, Department of Astronomy, University of Science and Technology of China, Hefei 230026, China}
\affiliation{School of Astronomy and Space Science, University of Science and Technology of China, Hefei 230026, China}

\author[0000-0002-4052-2394]{Kotaro Kohno}
\affiliation{Institute of Astronomy, The University of Tokyo, Osawa 2-21-1, Mitaka, Tokyo 181-0015, Japan}

\author[0000-0002-6588-1174]{Min Bao}
\affiliation{School of Physics and Technology, Nanjing Normal University, Nanjing 210023, China}
\affiliation{Department of Astronomy, Nanjing University, Nanjing 210093, China}
\affiliation{Institute of Astronomy, The University of Tokyo, Osawa 2-21-1, Mitaka, Tokyo 181-0015, Japan}

\author[0000-0003-3932-0952]{Kana Morokuma-Matsui}
\affiliation{Institute of Astronomy, The University of Tokyo, Osawa 2-21-1, Mitaka, Tokyo 181-0015, Japan}

\author[0000-0002-7660-2273]{Xu Kong}
\affiliation{CAS Key Laboratory for Research in Galaxies and Cosmology, Department of Astronomy, University of Science and Technology of China, Hefei 230026, China}
\affiliation{School of Astronomy and Space Science, University of Science and Technology of China, Hefei 230026, China}
\affiliation{Frontiers Science Center for Planetary Exploration and Emerging Technologies, University of Science and Technology of China, Hefei, Anhui, 230026, China}

\author[0000-0003-2682-473X]{Xiaoyang Chen}
\affiliation{National Astronomical Observatory of Japan, National Institutes of Natural Sciences, 2-21-1 Osawa, Mitaka, Tokyo 181-8588, Japan}


\begin{abstract}

High-resolution observations of ionized and molecular gas in the nuclear regions of galaxies are indispensable for delineating the interplay of star formation, gaseous inflows, stellar radiation, and feedback processes. Combining our new ALMA band 3 mapping and archival VLT/MUSE data, we present a spatially resolved analysis of molecular and ionized gas in the central 5.4 Kpc region of NGC 1365.  
We find the star formation rate/efficiency (SFR/SFE) in the inner circumnuclear ring is about 0.4/1.1 dex higher than in the outer regions.  At a linear resolution of 180 pc, we obtain a super-linear Kennicutt-Schmidt law, demonstrating a steeper slope (1.96$\pm$0.14) than previous results presumably based on lower-resolution observations.
Compared to the northeastern counterpart, the southwestern dust lane shows lower SFE, but denser molecular gas, and larger virial parameters. This is consistent with an interpretation of negative feedback from AGN and/or starburst, in the sense that the radiation/winds can heat and interact with the molecular gas even in relatively dense regions.  After subtracting the circular motion component of the molecular gas and the stellar rotation, we detect two prominent non-circular motion components of molecular and ionized hydrogen gas, reaching a line-of-sight velocity of up to 100 km/s. We conclude that the winds or shocked gas from the central AGN may expel the low-density molecular gas and diffuse ionized gas on the surface of the rotating disk.
\end{abstract}


\section{Introduction} \label{sec:intro}

The star formation mechanism in galactic environments is of fundamental importance to understand the formation and evolution of galaxies, such as the gas depletion, the metal enrichment, and the accumulation of stellar mass. Meanwhile, star formation results from the interplay of a series of perplexing processes and interactions, including the geometrical, dynamical, and chemical aspects of the interstellar medium (ISM), as well as feedback from young stellar objects, supernovae and the active galactic nuclei (AGN).

In barred spiral galaxies, star formation activities often behave differently in the spiral arms, interarm, and bar regions \citep[e.g.,][]{Momose2010, Foyle2010, Dobbs2014}. Optical and sub/millimeter observations show that young stars and molecular gas clouds are concentrated in the spiral arms, harboring higher star formation rate (SFR) densities and molecular gas densities. The cloud-cloud collisions in spiral arms and the gravitational collapse in giant molecular clouds (GMCs) are suggested to trigger star formation \citep{Elmegreen1983, Dobbs2008, Jeffreson2018, Elmegreen2019}. However, some observations \citep[e.g.,][]{Foyle2010,Eden2012,Eden2015,Kreckel2016} show that the difference of star formation efficiency (SFE) between arm and interarm regions is insignificant. In nuclear regions, star formation can also be affected by outflows and radio jets from the central AGN, for which two opposite scenarios have been proposed. On one hand, the energy from AGN may prevent the gas from cooling \citep{Fabian1994}, and massive and powerful outflows will sweep out gas from their host galaxies \citep{Fabian2012, Cheung2016May, Harrison2018Feb}, therefore the star formation will be suppressed (i.e., negative feedback). On the other hand, star formation in some high-density regions of the host galaxy will be triggered or enhanced by the outflows, because cold gas is compressed both in the galactic disk and in outflow regions \citep[i.e., positive feedback,][]{Silk2013, Cresci2015, Maiolino2017Mar, Gallagher2019}. Delineating the impact of AGN feedback on star formation is a long-standing challenge. Recently, using the VLT/MUSE and the Atacama Large Millimeter/submillimeter Array (ALMA) data, \cite{Shin2019} present a spatially resolved analysis of ionized and molecular gas, reporting both negative and positive feedback in the nearby Seyfert 2 galaxy NGC 5728.

An empirical scaling relation known as the Kennicutt-Schmidt law (or the K-S law) that links the surface density of SFR ($\Sigma_{\rm SFR}$) and molecular gas ($\Sigma_{\rm H_2}$), formulated as a single power law: $\Sigma_{\rm SFR} \propto \Sigma_{\rm H_2}^N$, has been well-established \citep{Schmidt1959,Kennicutt1998}. Thanks to the development of relevant instrumentation, extensive works \citep{Kennicutt2007, Bigiel2008, Momose2010, Liu2011, Momose2013Jul, Xu2015, Azeez2016, Wilson2019} have been undertaken, with a focus on the spatially resolved K-S law by measuring the SFR and gas properties at sub-Kpc scales. \cite{Bigiel2008} find the slope ($N$) of K-S law is nearly 1.0 at 750 pc resolution, while \cite{Kennicutt2007} obtained the $N \sim$ 1.56 for spiral galaxy M51 at 520 pc scales, where the linear vs. super-linear discrepancy is quantitatively explained by \cite{Liu2011}. For spiral galaxies, \cite{Momose2010} find the SFEs in arm regions to be twice of those in the bar regions, and the K-S law appear to break down at 250 pc resolution. \cite{Liu2011} and \cite{Momose2013Jul} remove the local diffuse emission in the $\ha$ and mid-infrared images, achieving a super-linear K-S law. In M51a and NGC 3521, \cite{Liu2011} also derive the variation of the slope as a function of linear resolution (250 pc – 1 Kpc) and find that the slope increases monotonically with increasing resolution. 

The wide variety of K-S laws found in different populations of galaxies using different tracers and spatial resolutions may be interpreted in terms of two theoretical frameworks of star formation in molecular gas. One is the density threshold model \citep[e.g., ][]{Gao2004,Wu2005,Lada2010,Lada2012,Evans2014}, which suggests that the fraction of dense gas in molecular clouds is a key factor and thus a linear correlation between SFRs and dense gas surface densities is found \citep[$\Sigma_{\rm gas} \ga 116 \ M_{\odot} \ \rm pc^2$, ][]{Lada2010}. The other is a turbulent model \citep[e.g., ][]{Krumholz2005,Krumholz2007,Federrath2012}, which is based on the integrals over the lognormal distribution of turbulent gas. \cite{Krumholz2007} find that the slope of K-S law is dependent on the choice of molecular line tracers, such as CO(1-0), HCO$^+$(1-0) and HCN(1-0), because different transition lines trace regions of different densities. \cite{Federrath2012} conclude that SFR in molecular clouds is controlled by the interstellar turbulence and the magnetic fields. Recently, some studies \citep[e.g., ][]{GarciaBurillo2012,Usero2015,Querejeta2019a,Genzel2020} find that the density threshold model is partly incompatible with the observed SFEs for dense gas, and suggest that the dynamical effects in local and global environments are to be taken into account as well.

In the nuclear Kpc region, \cite{Xu2015} reported the breaking down of K-S law at 100 pc scale in NGC 1614, in which the higher SFEs are probably triggered by the feedback of the central starburst. However, when focusing on the peaks of molecular and ionized gas at smaller scales, \cite{Kruijssen2019May} reported that SFE has two different branches, implying that the star formation in galactic GMCs is fast and inefficient because of the rapid feedback from radiation and stellar winds. The complication of the relation between star formation and molecular/ionized ISM is evident in spiral galaxies, especially in the circumnuclear Kpc region when a central AGN is present. High-resolution mapping of ionized and molecular gas in the nuclear regions are indispensable for unraveling the entanglements between star formation, gas inflows in the spiral arm, radiation, and outflows from AGN.

NGC 1365 is a nearby ($z$ = 0.0054, at a distance of 21.2 Mpc, where 1$\arcsec\sim$ 90 pc) archetype barred spiral (SBb(s)) Seyfert 1.8 galaxy with a stellar mass of $\sim 3.6 \times 10^{11} \ M_{\odot}$, and a member of the Fornax cluster \citep{Lindblad1999}. It harbors a low luminosity AGN with $L_{\rm bol} \sim 2 \times 10^{43} \ {\rm erg \ s^{-1}}$, exhibiting a biconical outflow and intense star formation in its central regions \citep[e.g.,][]{Galliano2005, Sakamoto2007, Galliano2008, Elmegreen2009, Wang2009, Galliano2012, Venturi2018a}. The position of the central AGN that we adopt is $\alpha = 03^{\rm h}33^{\rm m}36.35^{\rm s}$, $\delta=-36^{\circ}08{\arcmin}25.8\arcsec$, the systemic velocity, position angle (PA), and inclination angle are taken to be 1618 km/s, 220$^\circ$, and 40$^\circ$, respectively, following the parameter setting in \cite{Sakamoto2007}. A pair of dark dust lanes are located in front of the nuclear region and partially obscure the nucleus. Star formation and young massive star clusters are predominantly distributed in an elongated circumnuclear ring. However, the impacts of strong outflows from AGN or stellar winds from starburst on the gas consumption and star formation in NGC 1365 remain unexplored. In this work, with the facilitation of our new high-resolution sub/millimeter ALMA mapping and archival optical VLT/MUSE data, we investigate the connection between star formation, radiation, and inflows/outflows at sub-Kpc scales in the central $\sim 1^{'} \times 1^{'}$ (5.4 Kpc $\times$ 5.4 Kpc) region of this galaxy. 

This paper is organized as follows. In Section \ref{sec:data}, we describe the observations and data reduction. The properties of the molecular and ionized gas are analyzed in Section \ref{sec:ana}. The main results and discussion are presented in Section \ref{sec:resu} and \ref{sec:dis}, respectively, along with a summary in Section \ref{sec:sum}. We adopt a flat $\Lambda$CDM cosmology model throughout this work, $\Omega_\Lambda=0.7$, $\Omega_{\rm m}=0.3$, and $H_0=70$ km s$^{-1}$ Mpc$^{-1}$.

\section{Observations and data reduction}
\label{sec:data}

\begin{figure*}[t]
    \center
    \includegraphics[width=0.48\textwidth]{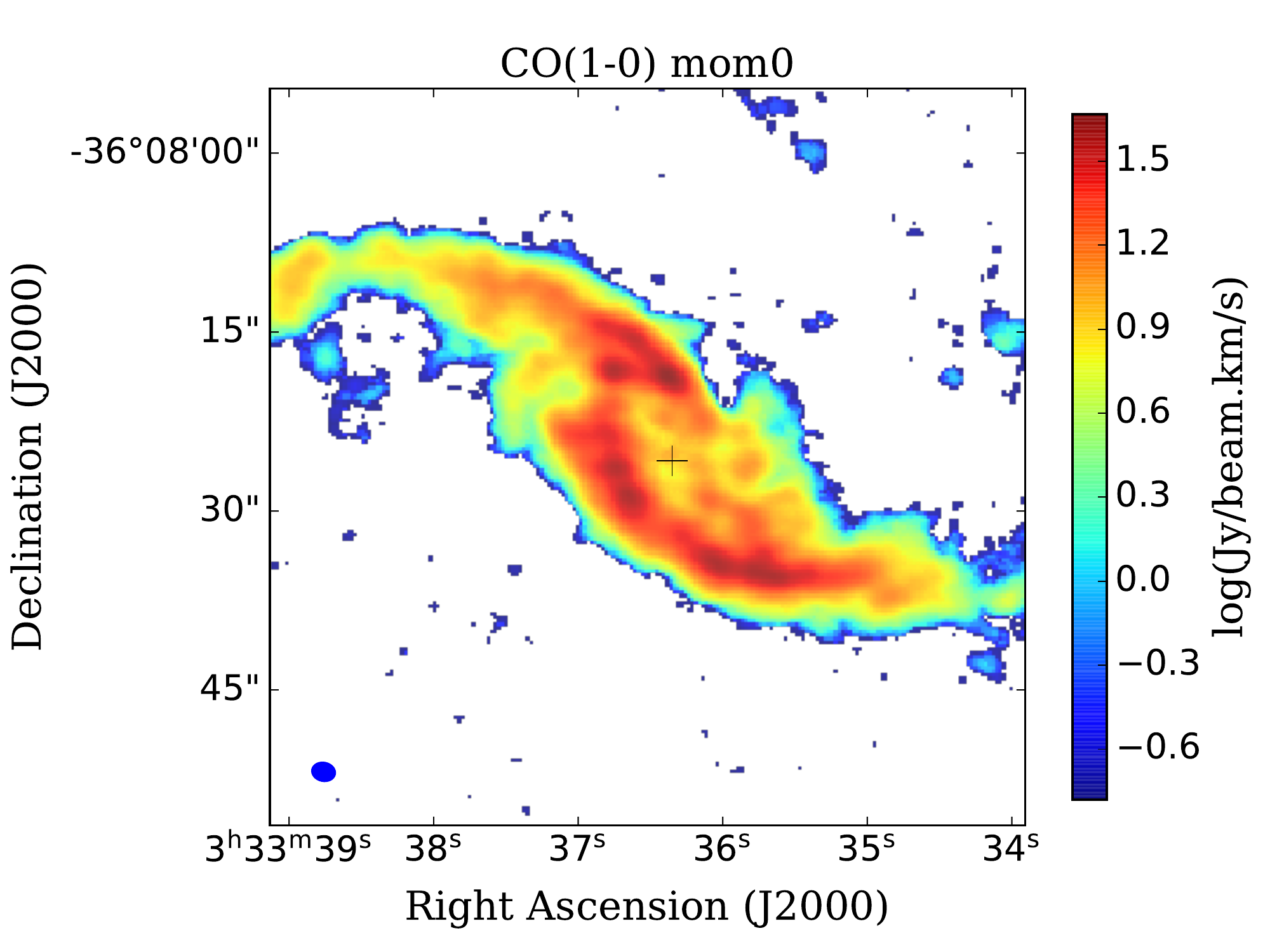}
    \includegraphics[width=0.48\textwidth]{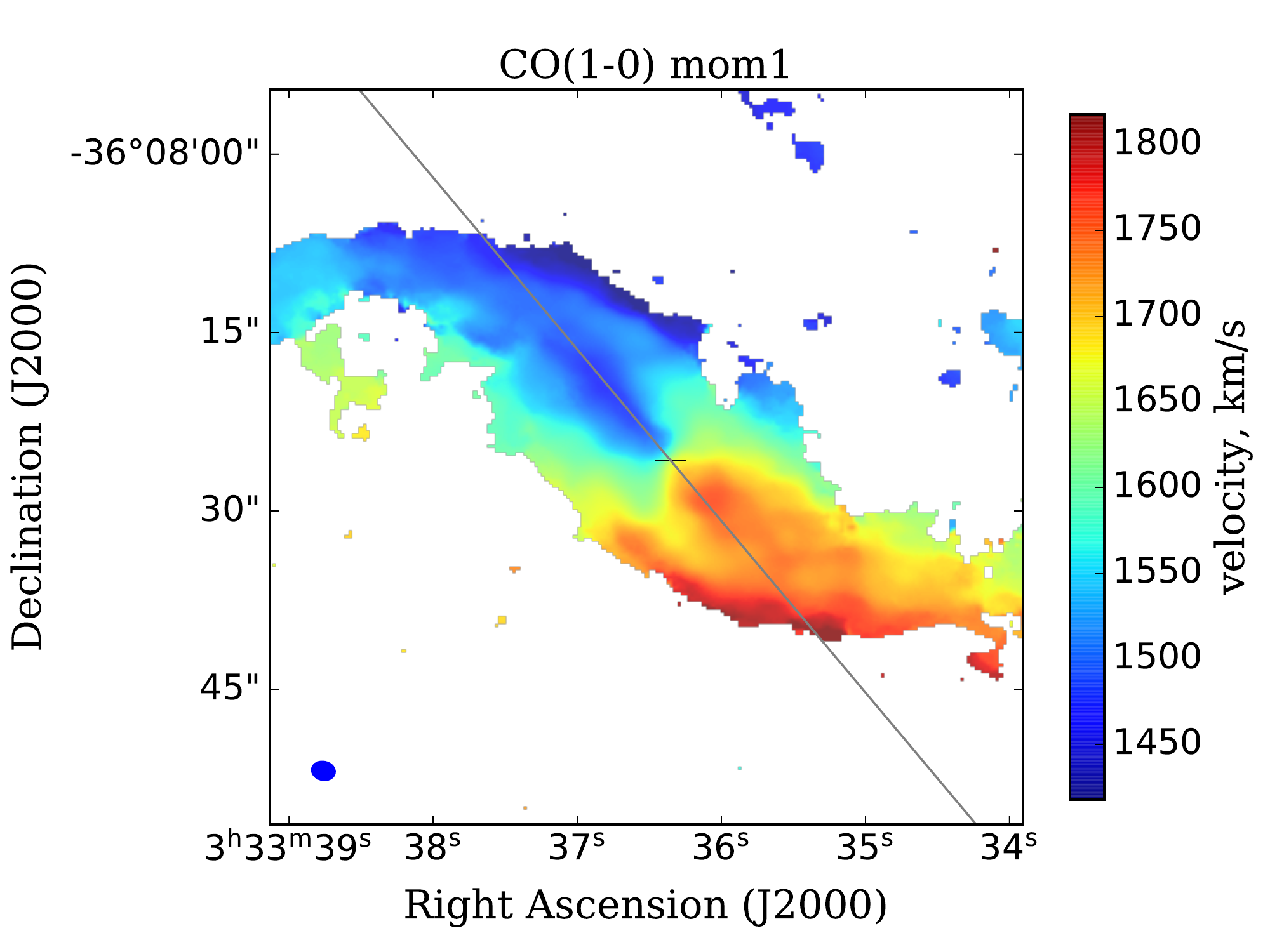}
    
    \includegraphics[width=0.48\textwidth]{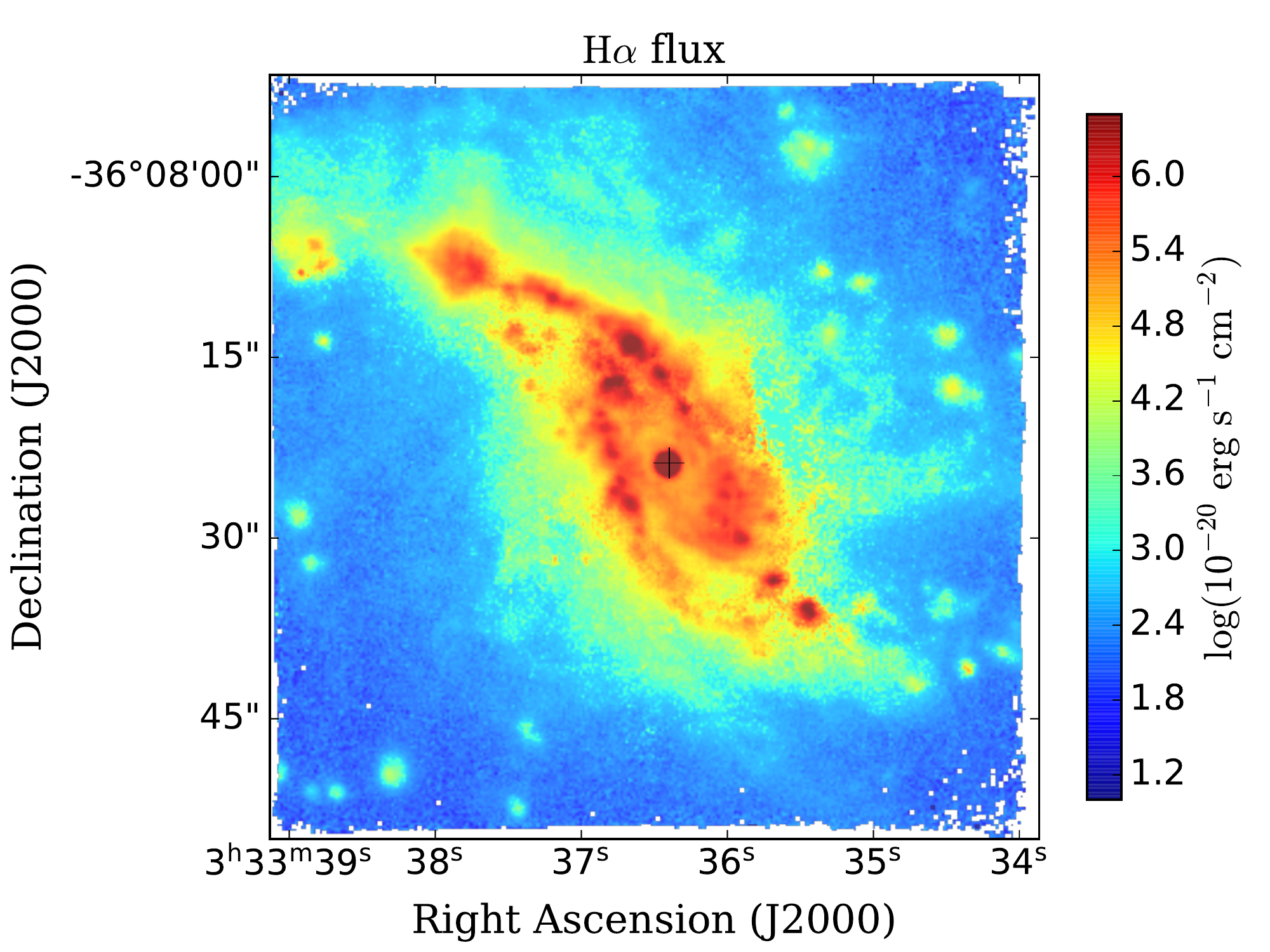}
    \includegraphics[width=0.48\textwidth]{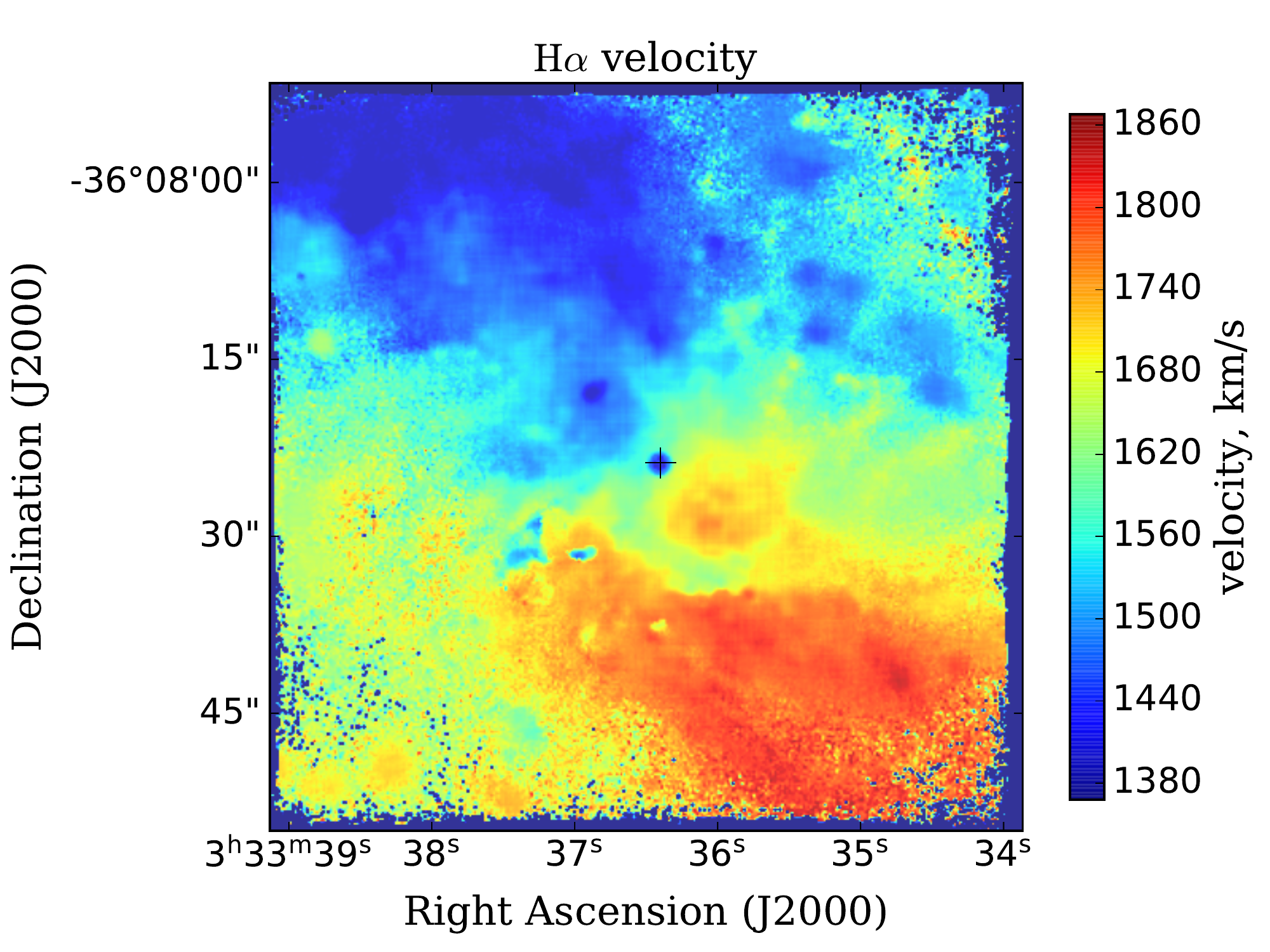}
    
    \includegraphics[width=0.48\textwidth]{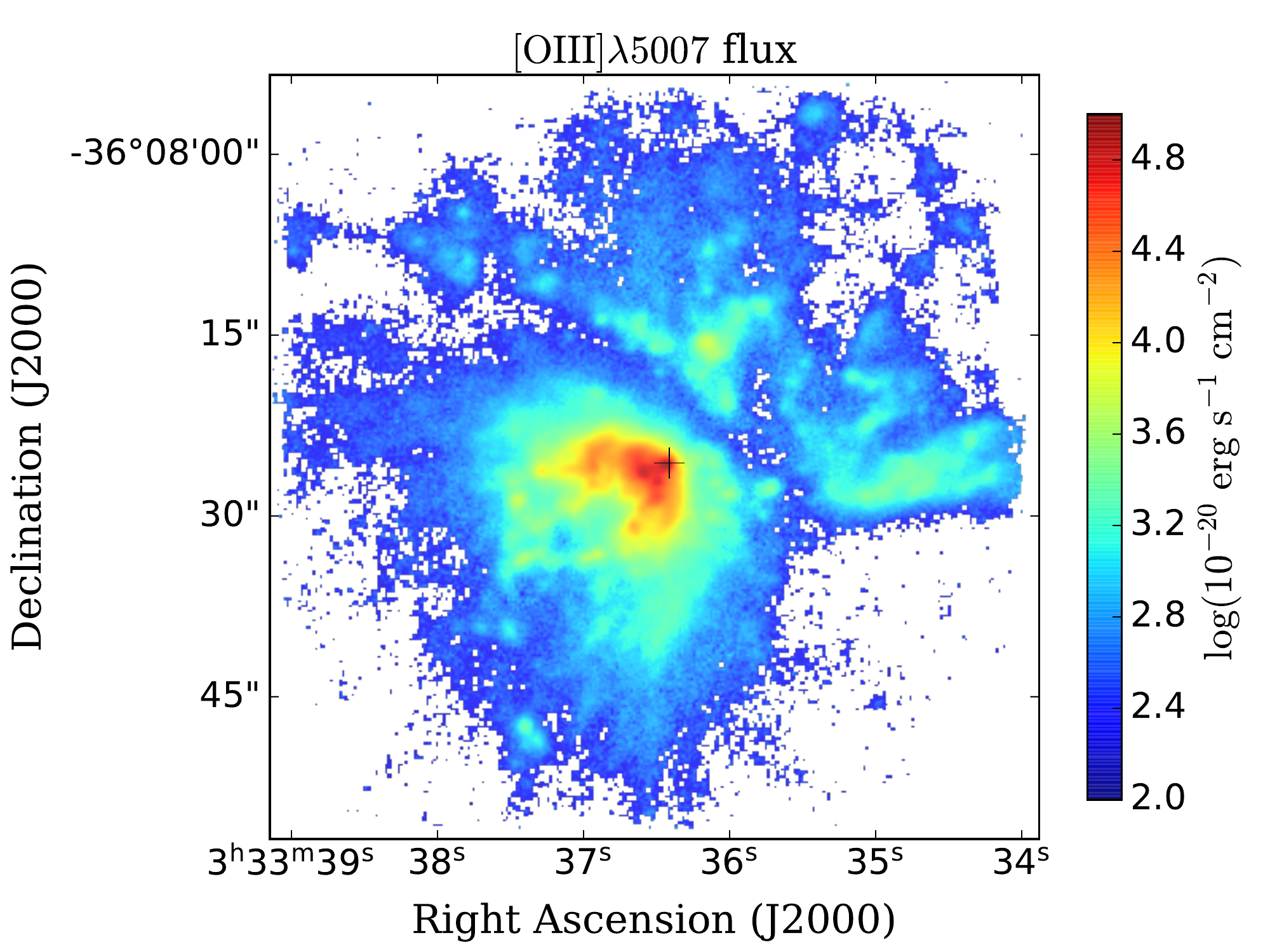}
    \includegraphics[width=0.48\textwidth]{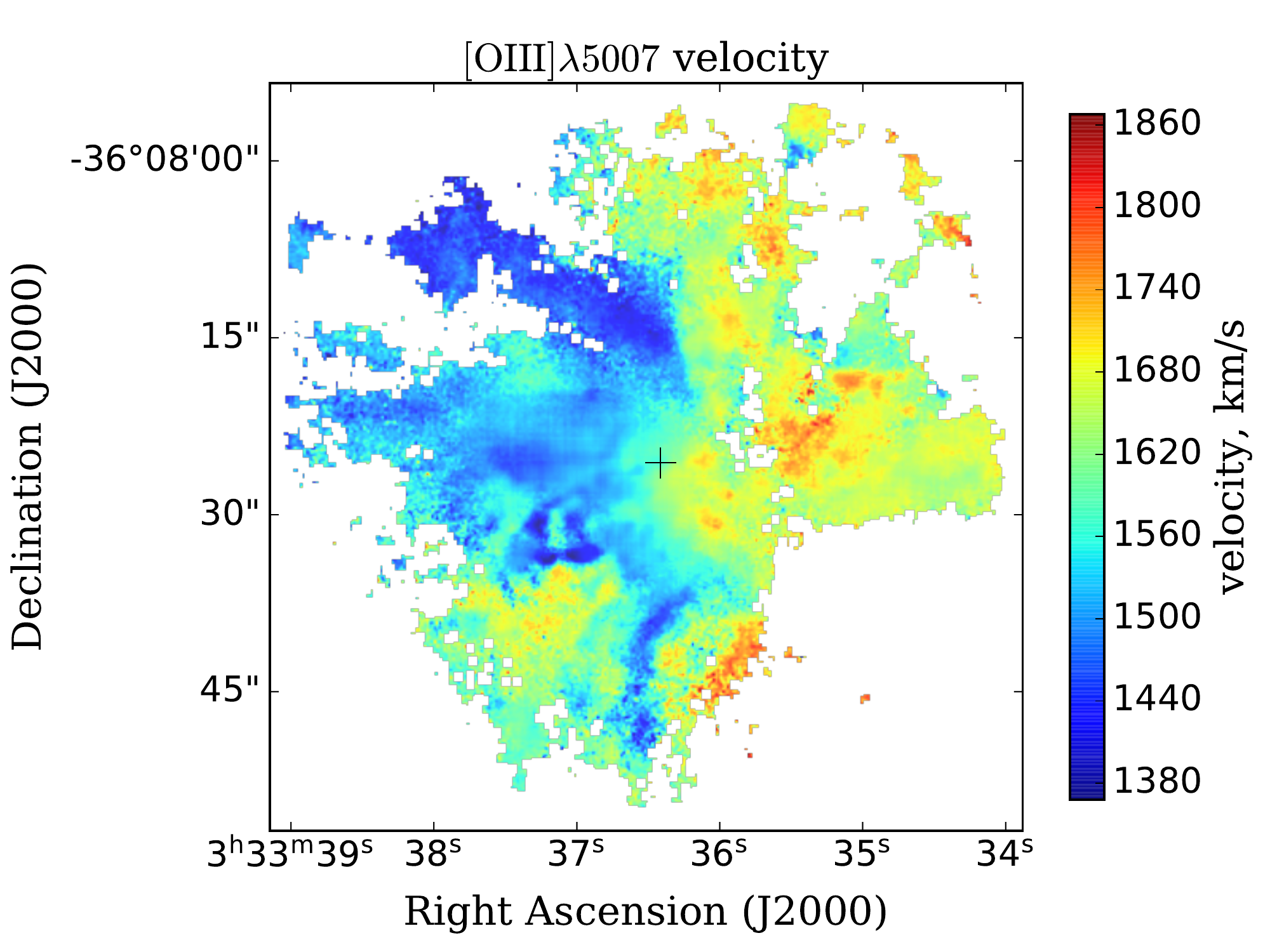}
    \caption{$Upper$: The integrated flux (moment 0, $left$) and mean velovity (moment 1, $right$) maps of CO(1--0) emission lines within the FoV of VLT/MUSE. The gray solid line represents the major axes, the systemic velocity is adopted as 1618 km/s. The threshold values in CO(1--0) maps are selected as 4$\sigma$. $Middle$: The integrated flux and velocity maps of attenuation corrected $\ha$ emission. $Bottom$: The integrated flux and velocity maps of $\oiii\lambda5007$ emission. The black crosses represent the central AGN position. }
    \label{fig:alma_muse_data}
\end{figure*}

\begin{deluxetable*}{lllllll}
\tablecaption{The observations for NGC 1365 \label{tab:obs}}
\tabletypesize{\scriptsize}  
\tablehead{
\colhead{Telescope} & \colhead{Project ID} & \colhead{PI} & \colhead{Wave/Frequency} &
\colhead{Resolution/Beam size} & \colhead{FoV} & \colhead{RMS} }
\startdata
VLT/MUSE         & 094.B-0321(A)  & Marconi & 4750-9352${\Ang}$ & 0.76\arcsec & 63.8\arcsec $\times$ 63.4\arcsec & {} \\
ALMA 12m+7m+TP  & 2015.1.01135.S, 2017.1.00129.S & Egusa; Morokuma & 99-115 GHz &  $1.92\arcsec \times 1.51\arcsec$ &  8\arcmin $\times$ 10\arcmin & 8.3 mJy/beam (5 km/s)
\enddata
\end{deluxetable*}

\subsection{ALMA data}
\label{subsec:alma}

NGC 1365 has been observed multiple times with ALMA in previous cycles. In this work, we map it in the CO (J=1-0) line, a cold molecular gas tracer, that falls in band 3 (Table \ref{tab:obs}). CO data of higher rotational transitions are also available in the ALMA archive, but we prefer to avoid complications in gas mass determination caused by heterogeneous excitation conditions. In band 3, the 12m-array observing campaign was undertaken on March 20th, 2016 (ID: 2015.1.01135.S, PI: Egusa). The integration time is about 1 minute per mosaic point (135 points in total). The correlator with a bandwidth of 468.7 MHz in spectral mode is adjusted to a central frequency of $\sim$ 115.27 GHz, so that the CO(J=1–0) line is covered. The 7m-array campaign was executed on Nov. 30th, 2017 (ID: 2017.1.00129.S, PI: Morokuma), with an integration time of 30 minutes, and the CO(J=1–0) emission line was also covered. The total power (TP) observation took place on Oct. 26th, 2019, with an integration time of 66,508 seconds. Standard calibrations were performed using the Common Astronomy Software Applications (CASA) package \citep[version 5.5,][]{McMullin2007}. The 12m and 7m data were imaged together with the CASA task \textit{tclean} using the briggs weighting with a robustness of 0.5 and with a threshold of 8.3 mJy for a channel width of 5 km/s. The cleaned data were then primary-beam corrected and combined with the TP data using the CASA task \textit{feather}. The synthesized beam size and RMS are listed in Table \ref{tab:obs}. CO moment maps were created by applying a 4$\sigma$ threshold. The moment 0 and 1 maps of CO(1–0) are presented in the top row of Figure \ref{fig:alma_muse_data}.

\subsection{Archival VLT/MUSE data}
\label{subsec:muse}

NGC 1365 was observed with the VLT/MUSE \citep{Bacon2010} on Oct. 12th, 2014, as a part of the Measuring Active Galactic Nuclei Under MUSE Microscope survey \citep[a.k.a MAGNUM, ID: 094.B-0321(A), PI: A. Marconi, ][]{Venturi2018a}. A fully reduced datacube is available on the ESO archive website \footnote{http://archive.eso.org/scienceportal/home}, which we use for our analysis in this work. The field of view (FoV) is $63.8\arcsec \times 63.4\arcsec$, consisting of 319 $\times$ 317 spaxels (0.2$\arcsec$ per each), and is compared to the entire galaxy in \citet[][Figure 1]{Venturi2018a}. The average seeing during the observations was about 0.76$\arcsec$. Using the python package MPDAF\footnote{\url{https://mpdaf.readthedocs.io/en/latest/start.html}}\citep{Bacon2016}, the median value of FWHM of PSF at $4500 \rm \Ang < \lambda < 7000 \rm \Ang$ is about 0.8$\arcsec$. The spectral window is 4750 -- 9352$\rm \Ang$ and the channel width is 1.25$\rm \Ang$.

\section{data analysis}
\label{sec:ana}

\subsection{Molecular gas mass from CO(1--0)}
\label{subsec:almadata}

In order to compare the molecular gas to the ionized gas, we construct the CO(1--0) map within the FoV of the MUSE data, which are shown in Fig. \ref{fig:alma_muse_data}. Following \cite{Sakamoto2007}, we assume a $^{12}$CO to $\rm H_2$ conversion factor of  $X_{\rm CO} = \rm 0.5 \times 10^{20} cm^{-2} (K \ km \ s^{-1})^{-1}$ , a value commonly adopted for galactic centers and starburst nuclei. Note that adopting alternative $X_{\rm CO}$ values \citep[e.g.,][]{Bolatto2013} will not change the slope and scatter in the K-S law. Hence, the molecular gas mass can be estimated as

\begin{equation}
\begin{split}
M_{\rm H_2} = 1.0 \times 10^4 (\frac{S_{\rm CO}}{\rm Jy \ km \ s^{-1}}) (\frac{D}{\rm Mpc})^2 \\
 \times [\frac{X_{\rm CO}}{\rm 0.5 \times 10^{20} cm^{-2} (K \ km \ s^{-1})^{-1}}] \msun,
\end{split}
\end{equation}
where $D = 21.2$ Mpc is the distance of NGC 1365 \citep{Sakamoto2007}. We devide the molecular gas mass in each spaxel by its corresponding physical area (corrected by the inclination angle), to determine the molecular gas mass surface density $\Sigma_{\rm H_2}$. The lowest $\Sigma_{\rm H_2}$ value that we detect is about $2.7 \times 10^6 \ M_{\odot} \rm \ Kpc^{-2}$, which is adopted as 4$\sigma$ threshold.

\subsection{SFR and stellar mass from MUSE data}
\label{subsec:musedata}

We perform a series of spectral fitting analysis on the spectra in the MUSE datacube, whose spectral coverage is 4750 -- 8500$\rm \Ang$ with the primary optical emission lines covered (e.g., $\hb$, $\oiii\lambda\lambda4959,5007$, $\ha$, $\nii\lambda\lambda6548,6583$ and $\sii\lambda6717,6731$). With the optical emission lines masked, we employ the STARLIGHT routine \citep{CidFernandes2005} to recover the underlying stellar continuum. Assuming the initial mass function (IMF) from \cite{Chabrier2003}, we fit each spectrum to a combination of 45 single stellar populations (SSPs) from the \cite{Bruzual2003} model, which are distributed on three different metallicities (Z = 0.01, 0.02, and 0.05) and 15 stellar ages (1 Myr to 13 Gyr). We obtain the stellar mass within each spatial pixel from the SSP fitting results. When the signal-to-noise ratio (S/N) of the continuum is above 5, the uncertainty of the stellar mass is smaller than 0.11 dex \citep{Bruzual2003, CidFernandes2005}.

To optimally recover the fluxes of the strong emission lines ($\hb$, $\oiii\lambda\lambda4959,5007$, $\ha$), we fit them to multiple Gaussians using the IDL package MPFIT \citep{Markwardt2009}. The estimation of S/N for these emission lines is done following the method used in \cite{Ly2014} and \cite{Gao2018Dec}. The velocities of stellar and ionized gas are achievable from the absorption lines embedded in the continuum and from the strong emission lines, respectively. In Fig. \ref{fig:alma_muse_data}, we show the integrated flux and velocity maps for strong emission lines  $\ha$ and $\oiii\lambda5007$. To obtain extinction-corrected SFRs, we locate the regions in the $\ha$ and $\hb$ images with S/N larger than 5, and derive $\rm A_v$ with $\ha/\hb$ ratios, employing ``Case B'' recombination model and \cite{Calzetti2000} reddening formalism. Assuming the solar metallicity and \cite{Chabrier2003} IMF, SFR can be derived from the $\ha$ luminosity by ${\rm SFR}(\msun {\rm \ yr^{-1}}) = 4.4 \times 10^{-42} \times L_{\rm cor} \rm (\ha)(erg \ s^{-1})$ \citep{Kennicutt1998}. For stellar mass and SFR, we compute their surface densities, $\Sigma_{*}$ and $\Sigma_{\rm SFR}$, respectively, as we do for $\Sigma_{\rm H_2}$. 

\subsection{The spatially-resolved BPT diagram and AGN fraction}
\label{subsec:bpt}

\begin{figure*}[t]
\center
\includegraphics[width=0.4\textwidth]{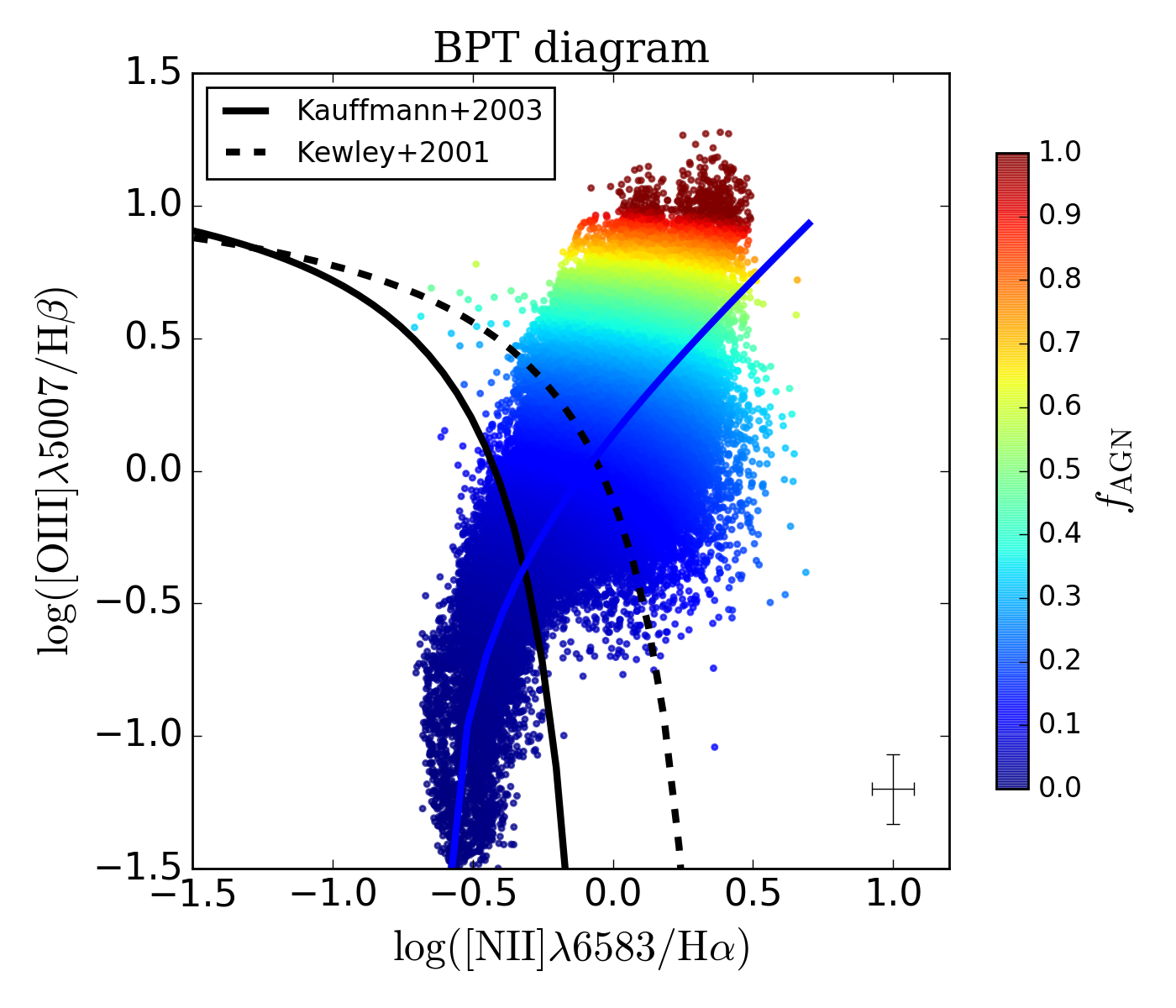}
\includegraphics[width=0.45\textwidth]{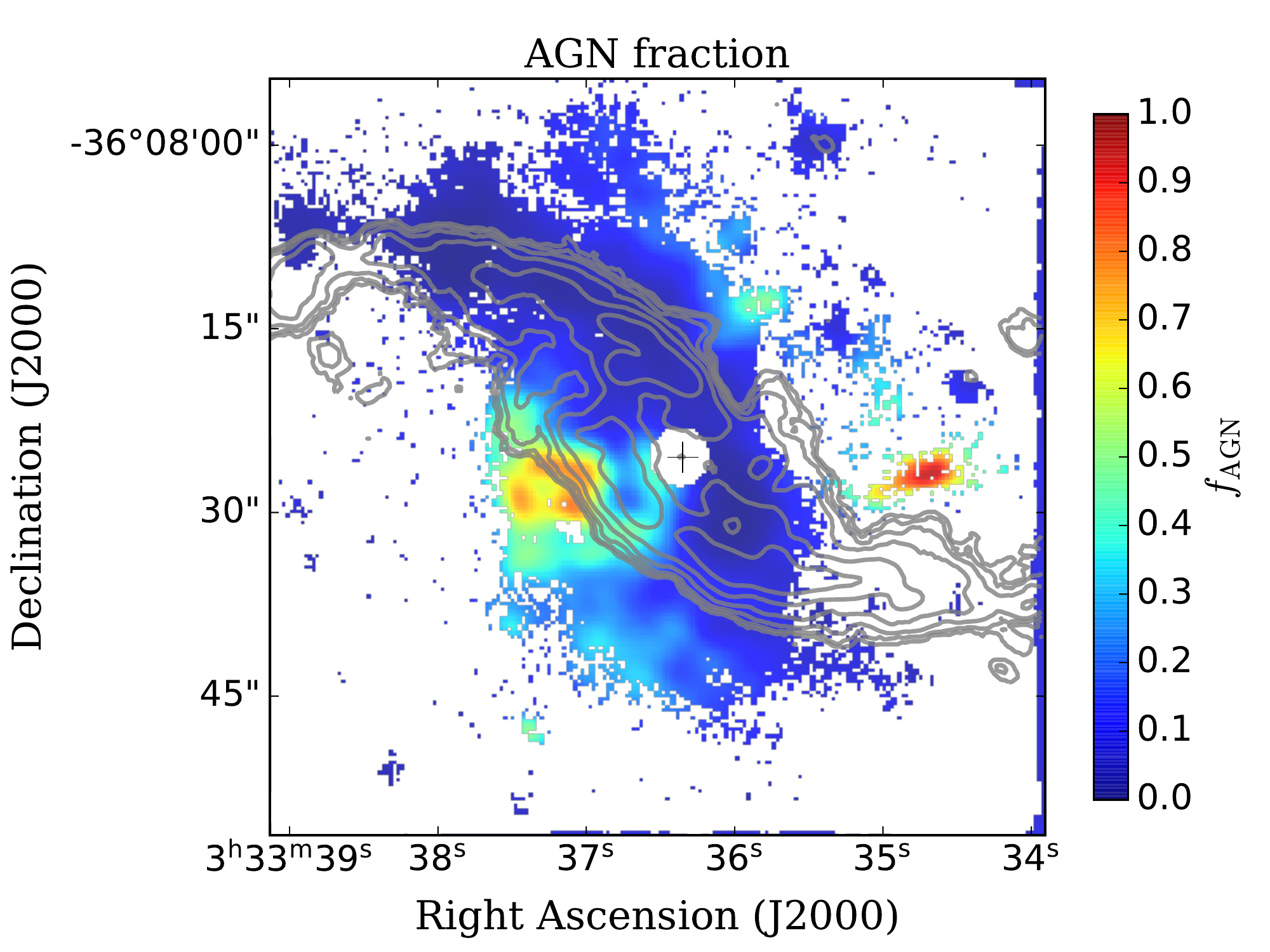}
\caption{$Left$: The pixel-based BPT diagram with $\nii$ diagnosis for NGC 1365. The color means the contribution of AGN, ranging from 0 to 100$\%$. Blue line denotes the mixing sequence between star forming region and AGN region. $Right$ panel: The spatial resolved AGN fraction map. The gray contours represent the CO(1-0) integrated intensity, ranging at  $(1/2)^n (n = 0-6)$ times of the largest flux. The black cross shows the central AGN position, and the central 2.4$\arcsec$ region is masked because of the contamination from broad line region. }
\label{fig:bpt}
\end{figure*}

We construct a pixel-based BPT diagram \citep{Baldwin1981, Kewley2001, Kauffmann2003} using the $\nii$ diagnosis to scrutinize the ionization state of the nuclear region of NGC 1365 in Figure \ref{fig:bpt} (left panel), similar to \citet[Figure 5]{Venturi2018a}. We derive the AGN contribution fraction ($f_{\rm AGN}$) for each spaxel following the method used in \cite{Davies2016} and \cite{Shin2019}. The AGN fraction spans a wide range (from 0$\%$ to 100$\%$, Figure 2).  As seen in the right panel of Fig. \ref{fig:bpt}, $f_{\rm AGN} < 20\%$ in the circumnuclear ring, consistent with previous studies \citep[e.g., ][]{Davies2016, Agostino2019, Shin2019}. In the biconical outflow, in contrast, $f_{\rm AGN} > 60\%$. We then correct SFR and $\Sigma_{\rm SFR}$ by quantifying the star-formation fraction in the $\ha$ luminosities. For comparison, \cite{Durre2018} report that the AGN fraction of the star-forming ring in NGC 5728 is about 40$\%$, which is caused by the determination of fractions in logarithmic space. Nevertheless, even if we assume the AGN fraction in the circumnuclear ring to be 40$\%$, the SFR distribution remains comparable to our results given in Section \ref{sec:resu}.

\section{Results}
\label{sec:resu}

\subsection{Star formation relations in nuclear region}
\label{subsec:SF}

\begin{figure*}[t!]
\center
\includegraphics[width=0.32\linewidth]{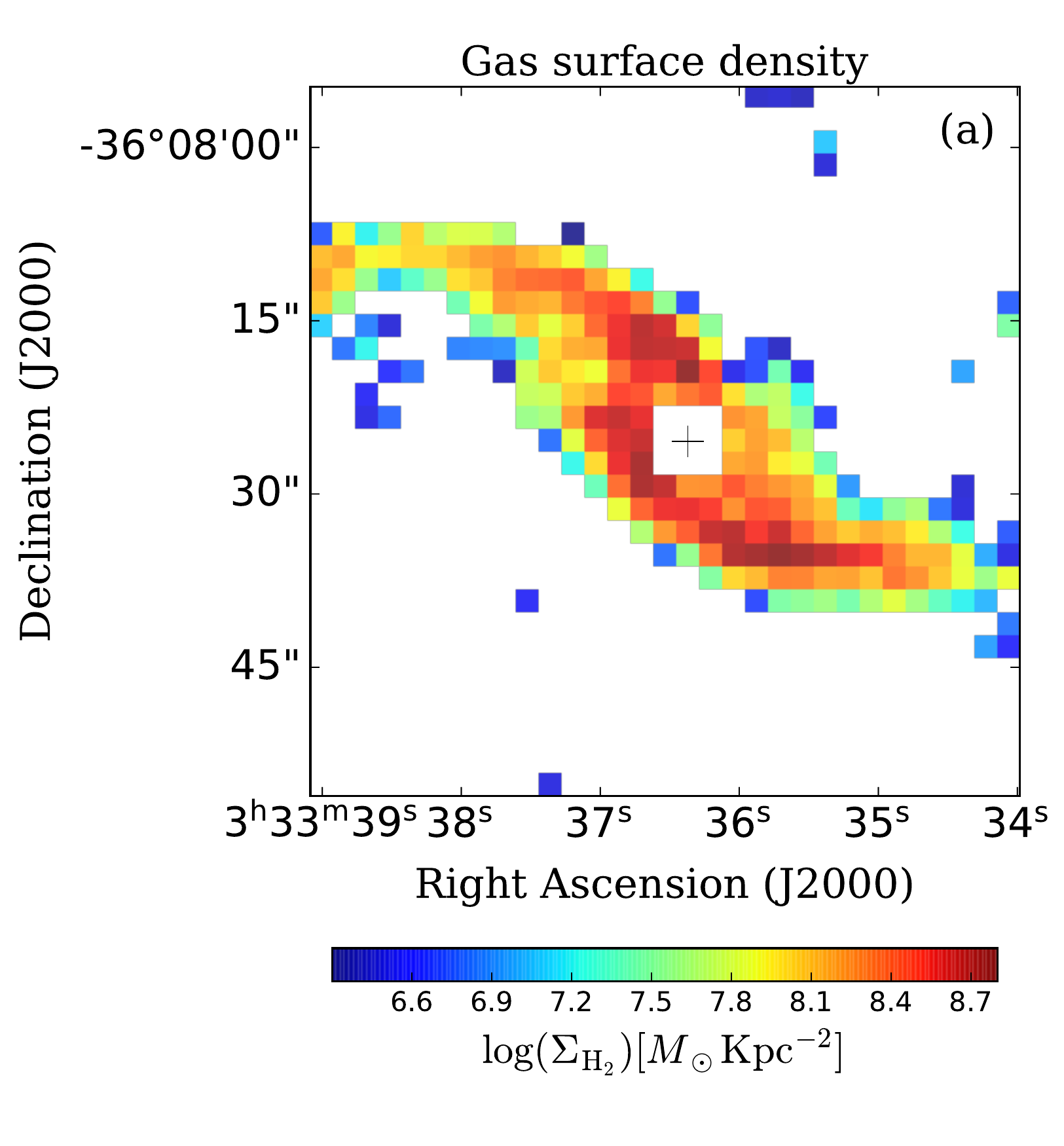}
\includegraphics[width=0.32\linewidth]{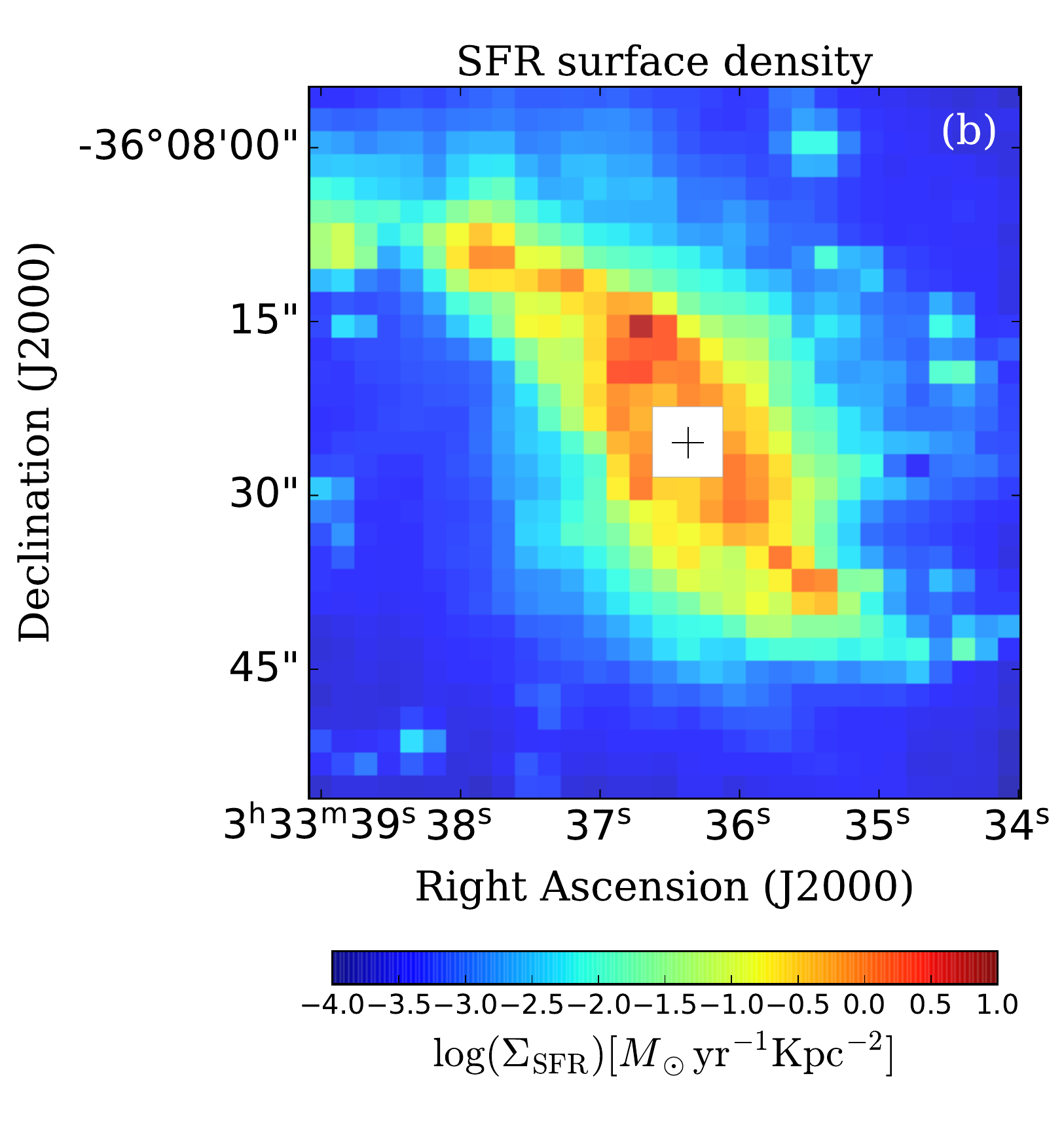}
\includegraphics[width=0.32\linewidth]{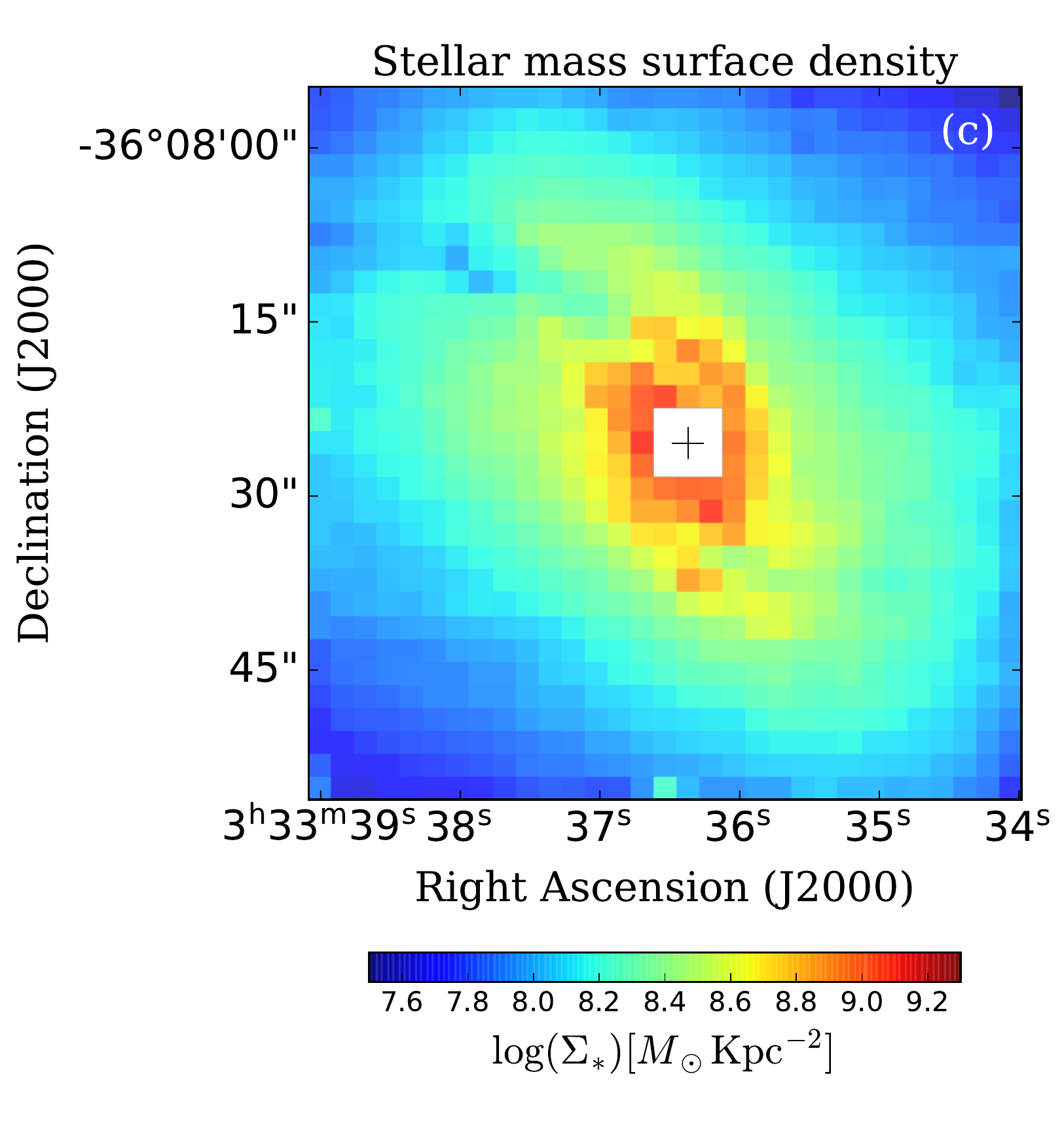}
\caption{$a,b,c$: The surface density maps of molecular gas mass, SFR and stellar mass at a same resolution of 2\arcsec ($\sim 180$ pc). We mask the central 9 pixels around the AGN position.}
\label{fig:gas_sfr_stellar_av_regrid}
\end{figure*}

\begin{figure*}[t!]
\center
\includegraphics[width=0.48\linewidth]{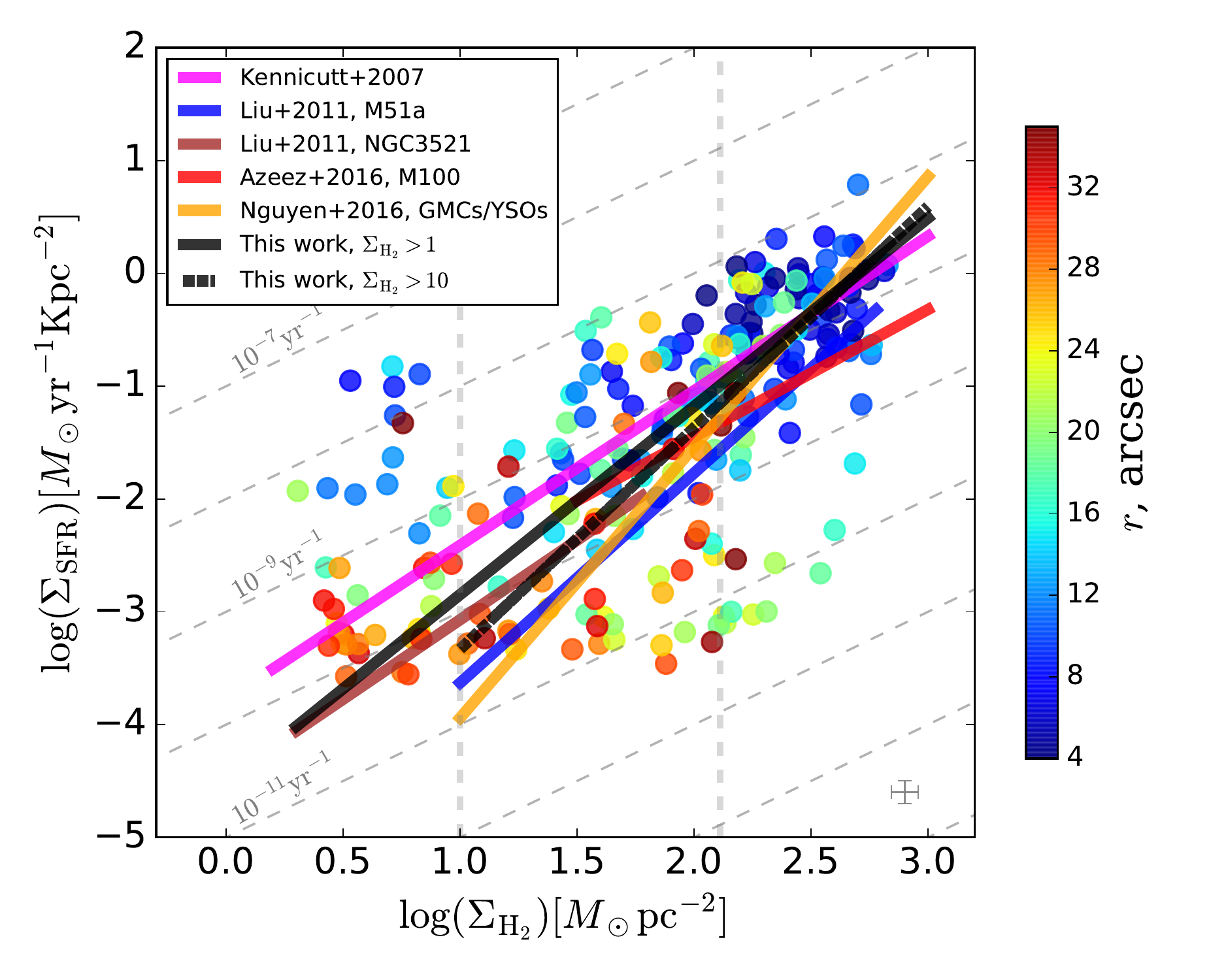}
\includegraphics[width=0.48\linewidth]{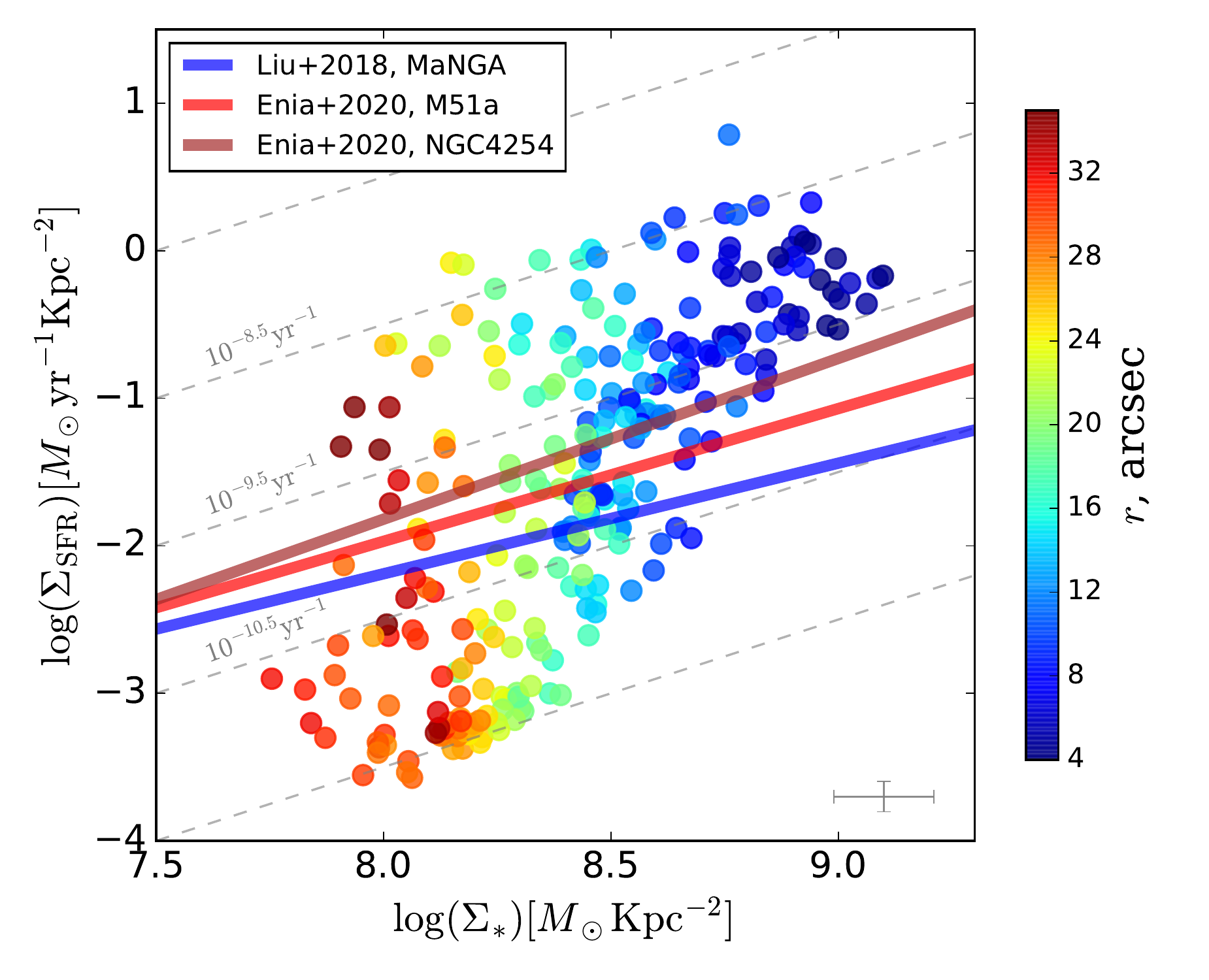}
\caption{$Left$: The SFR surface densities as a function of molecular gas surface densities, the resolved K-S law, colored by the distance ($r$) between its pixel to central AGN. The cyan, blue, brown, red and orange solid lines show the K-S laws in M51a \citep[500 pc, 250 pc, ][]{Kennicutt2007,Liu2011}, NGC 3521 \citep[250 pc, ][]{Liu2011}, M100 \citep[330 pc, ][]{Azeez2016} and GMCs/YSOs \citep{NguyenLuong2016}, respectively. The black solid and dashed line represent the best fitted relations for the data with $\Sigma_{\rm H_2} > 10 \ M_{\odot} \ \rm pc^{-2}$ and $\Sigma_{\rm H_2} > 1 \ M_{\odot} \ \rm pc^{-2}$, respectively. The gray dashed lines represent different SFE levels. The two vertical gray lines seperate the gas into low, intermediate and high density regimes \citep[Table 3, ][]{Kennicutt2012}. $Right$: The SFR surface densities versus the stellar mass surface densities, known as resloved main sequence relation. The blue, red and brown solid lines represent the resolved relations in MaNGA survey \citep{Liu2018}, M51a and NGC 4254 \citep{Enia2020}, with a resolution of 1-2 Kpc, 330 and 500 pc, respectively. The gray dashed lines represent different sSFR levels. The errorbars show the median values of determination uncertainties. }
\label{fig:ks_msr}
\end{figure*}

In this section, we analyze the spatially-resolved K-S relation, and the stellar mass – SFR (main sequence) relation in the nuclear region of NGC 1365. Regarding the fact that the beam size in our CO(1--0) data is 1.9\arcsec $\times$ 1.5\arcsec, a spatial resolution significantly lower than that of the SFR and stellar mass surface density maps, we regrid the latter maps to a resolution of 2\arcsec $\times$ 2\arcsec ($\sim$ 180 pc $\times$ 180 pc) to prevent from oversampling. To minimize the AGN contamination to the determination of SFR and stellar mass, we also mask the central nine pixels. In the $a, b, c$ panels of Fig. \ref{fig:gas_sfr_stellar_av_regrid}, we show the surface density maps of molecular gas mass, SFR and stellar mass at a resolution of $\sim 180$ pc. 

In the $Left$ panel of Fig. \ref{fig:ks_msr}, we plot the spatially-resolved molecular gas surface density vs. SFR density relation, where data points are color-coded by their galacto-centric distances ($r$). We also compare our analysis with the K-S laws derived in 5 previous investigations in Fig. \ref{fig:ks_msr}: (1) \cite{Kennicutt2007} at 500 pc scale in the M51a disk; (2, 3) \cite{Liu2011} in M51a and NGC 3521 at spatial resolutions ranging from 250 pc to $\sim$1 kpc; (4) \cite{Azeez2016} in the central region of M100 (NGC 4321) at 330 pc resolution; (5) \cite{NguyenLuong2016} using the young stellar objects (YSOs) at 30 pc resolution in the dense GMCs of the Galaxy. The gray dashed lines represent different SFE levels ($10^{-7}$, $10^{-9}$ and $10^{-11} \ \rm yr^{-1}$). The two vertical gray lines denote low, intermediate, and high-density regimes, corresponding to the regimes with sparse, moderate, and concentrated star formation activities \citep[see Table 3, ][]{Kennicutt2012}. 

The Pearson and the Spearman coefficients between gas density and SFR density are about 0.67 and 0.72, respectively. Following \cite{Kennicutt2007}, the K-S relation can be expressed as
\begin{equation}
{\rm log}(\frac{\Sigma_{\rm SFR}}{\msun \ \rm yr^{-1} \ Kpc^{-2}}) =
N \ {\rm log}(\frac{\Sigma_{\rm H_2}}{\msun \ \rm pc^{-2}}) + A,
\end{equation}
where $N$ denotes the power-law index of the power law. We fit the $\Sigma_{\rm H_2}$ vs. $\Sigma_{\rm SFR}$ relation at $\Sigma_{\rm H_2} > 10 \ M_{\odot} \rm \ pc^{-2}$ and $\Sigma_{\rm H_2} > 1 \ M_{\odot} \rm \ pc^{-2}$ in sequence, based on an orthogonal distance regression algorithm ($scipy.odr$\footnote{https://docs.scipy.org/doc/scipy/reference/odr.html}), shown as the black dashed and solid lines, respectively. The best-fit results, compared to previous works, are listed in Table \ref{tab:ks}. The K-S law shows a steeper slope here than in M51a \citep{Kennicutt2007} and NGC 3521 \citep{Liu2011} when the applied threshold is $\Sigma_{\rm H_2} > 1 \ M_{\odot} \rm \ pc^{-2}$. Furthermore, if we restrict the measurement to $\Sigma_{\rm H_2} > 10 \ M_{\odot} \rm \ pc^{-2}$, the slope of K-S law is roughly consistent with in M51a \citep{Liu2011}, while is steeper than in M100 \citep{Azeez2016}. However, our slopes are flatter than the relation derived from the YSOs in the GMCs of the Galaxy. This is in line with the conclusion of \cite{Liu2011} that the spatial resolution affects the shape and slope of these star formation relations.

\begin{deluxetable*}{lcccl}[!]
\tablecaption{The comparison between different K-S laws \label{tab:ks} and our results.}
\tablehead{
\colhead{Galaxy} & \colhead{log10($\Sigma_{\rm H_2}/M_{\odot} \rm \ pc^{2}$)} & \colhead{Resolution (pc)} & \colhead{$N$} & \colhead{Reference}  }
\startdata
M51a  & 0 -- 3.0  & $>$ 500 & 1.37$\pm$0.03 & \cite{Kennicutt2007} \\
M51a  & 1 -- 2.5 & 250 & 1.86$\pm$0.03 & \cite{Liu2011} \\
NGC 3521  & 0 -- 1.5 & 250 & 1.41$\pm$0.06 & \cite{Liu2011} \\
M100  & 2.2 -- 3.1 & 330 & 1.13$\pm$0.20 & \cite{Azeez2016} \\
GMCs/YSOs  & 0 -- 4.5 & $<$ 30 & 2.40 & \cite{NguyenLuong2016} \\
NGC 1365 & 1 -- 3.0 & 180 & 1.96$\pm$0.14 & This work \\
NGC 1365 & 0 -- 3.0 & 180 & 1.67$\pm$0.10 & This work \\
\enddata
\end{deluxetable*}

The inner regions ($r < 9\arcsec$) are found to have about an SFE (SFR) 1.1 (0.4) dex higher than the outer regions, indicating that more intense star formation activities are occurring in the dust lane and the star-forming ring around the AGN. However, there exist dense regions with $\Sigma_{\rm H_2} > 100 \ M_{\odot} \rm \ pc^{-2}$ but with significantly lower SFE ($\sim 10^{-11} \ \rm yr^{-1}$), indicating suppression of star formation in a number of dense gas regions. In parallel, in the low-density regime, a number of regions show relatively high SFE ($\sim 10^{-8} \rm \ yr^{-1}$), which are scattered in both inner and outer locations, indicating remarkably enhanced star formation.

In the $Right$ panel of Fig. \ref{fig:ks_msr}, we present the stellar mass surface density – SFR surface density relation for these molecular regions, color-coded by their distances to the central AGN. The blue, red, and brown solid lines represent the resolved relations in star-forming galaxies from the MaNGA survey \citep{Liu2018}, M51a, and NGC 4254 \citep{Enia2020}, measured at a resolution of 1-2 Kpc, 330 and 500 pc, respectively. The gray dashed lines represent different specific SFR (sSFR $=$ SFR / $M_*$) levels, annotated with three values, $10^{-8.5}$, $10^{-9.5}$ and $10^{-10.5} \ \rm yr^{-1}$. We note that $\Sigma_{\rm SFR}$ increases rapidly with increasing $\Sigma_{*}$. The steeper slope indicates that the accumulation of stellar mass within the nuclear regions of NGC 1365 is significantly faster than normal star-forming galaxies and spiral galaxies, which might be caused by the enhancement of strong bars on star formation. High sSFR values ($> 10^{-9.5} \rm \ yr^{-1}$) are found both in the inner and outer regions, yet the majority of the outer regions possess low sSFR ($< 10^{-10.5} \rm \ yr^{-1}$). This result indicates the co-existence of enhancement and suppression of star formation activities in the nuclear region. However, the underestimated stellar mass as a result of the severe extinction in the central regions (cf. Fig. \ref{fig:SFE}, upper - right panel) might steepen the main sequence relation.

\subsection{Spatial distribution of SFE}
\label{subsec:sfe_distribution}

\begin{figure*}[t]
\center
\includegraphics[width=0.48\textwidth]{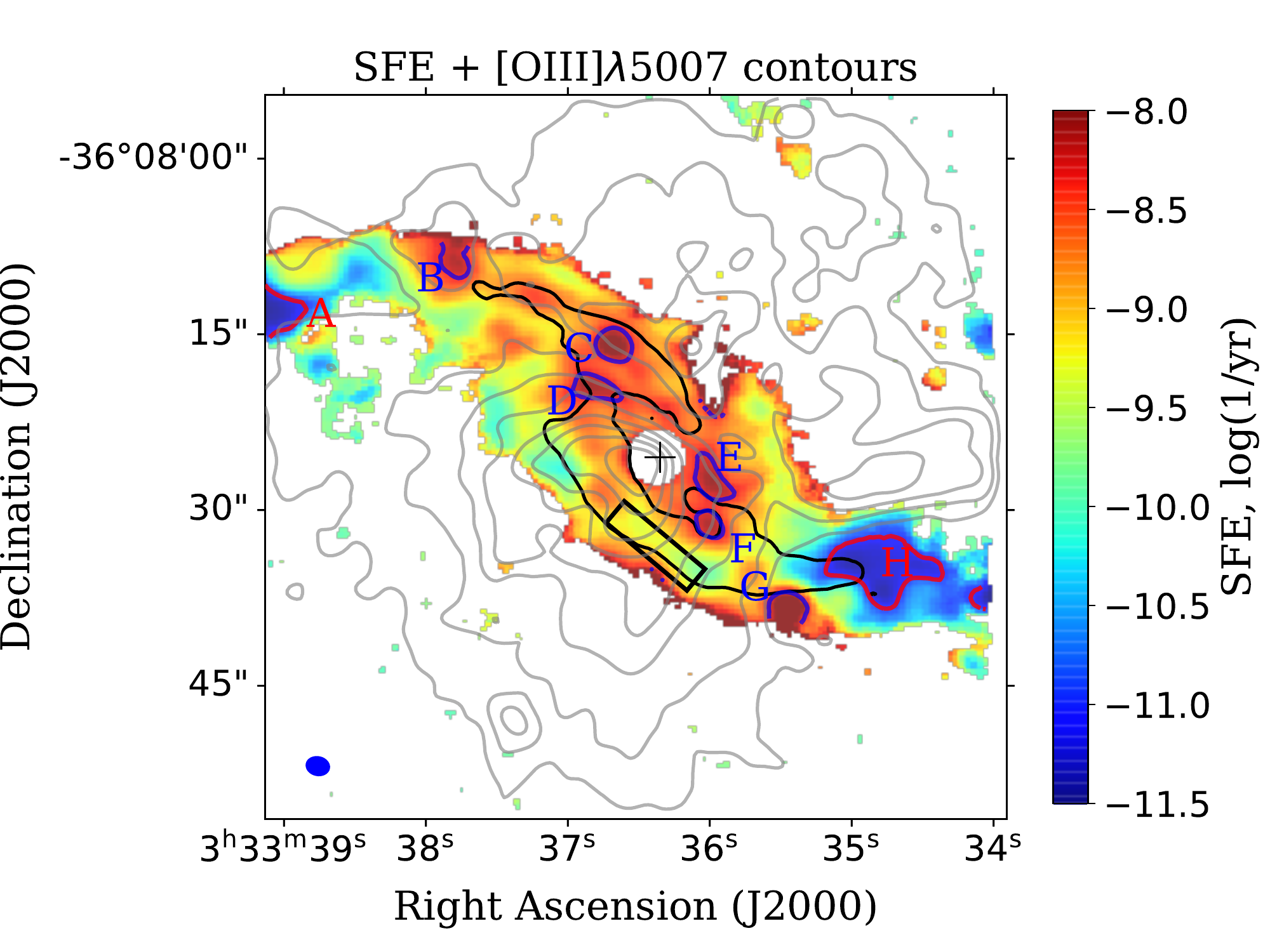}
\includegraphics[width=0.48\textwidth]{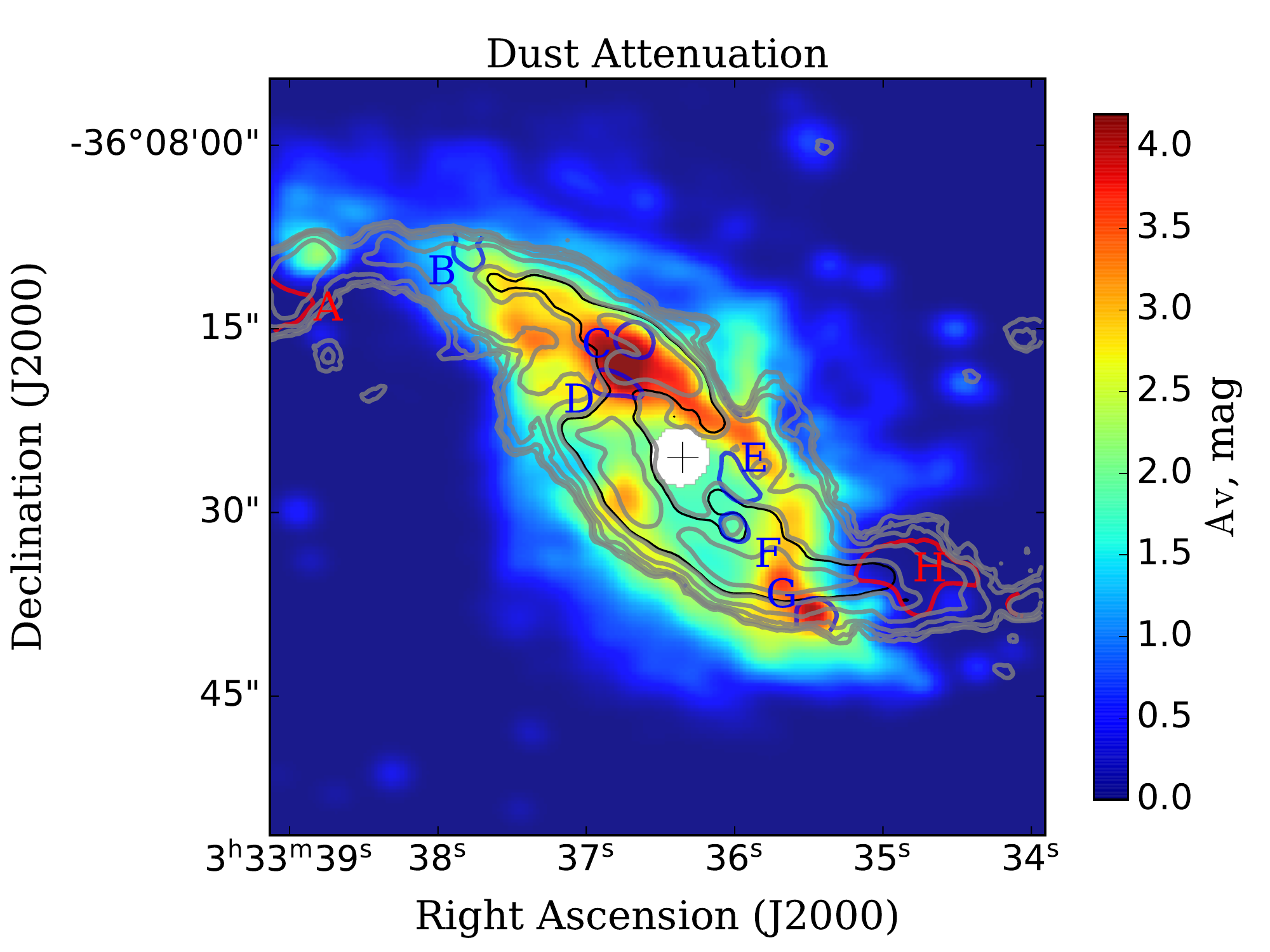}
\includegraphics[width=0.48\textwidth]{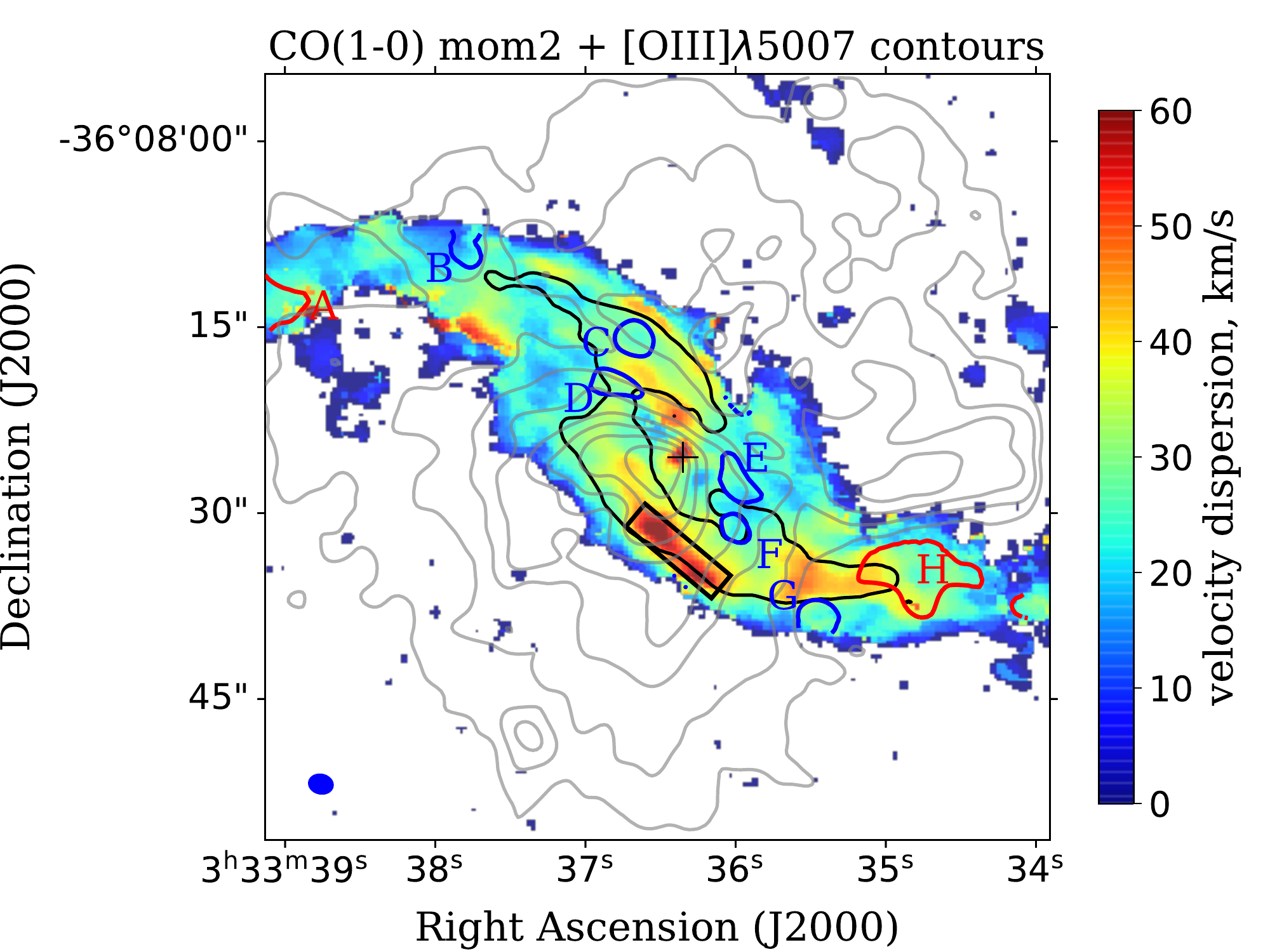}
\includegraphics[width=0.48\textwidth]{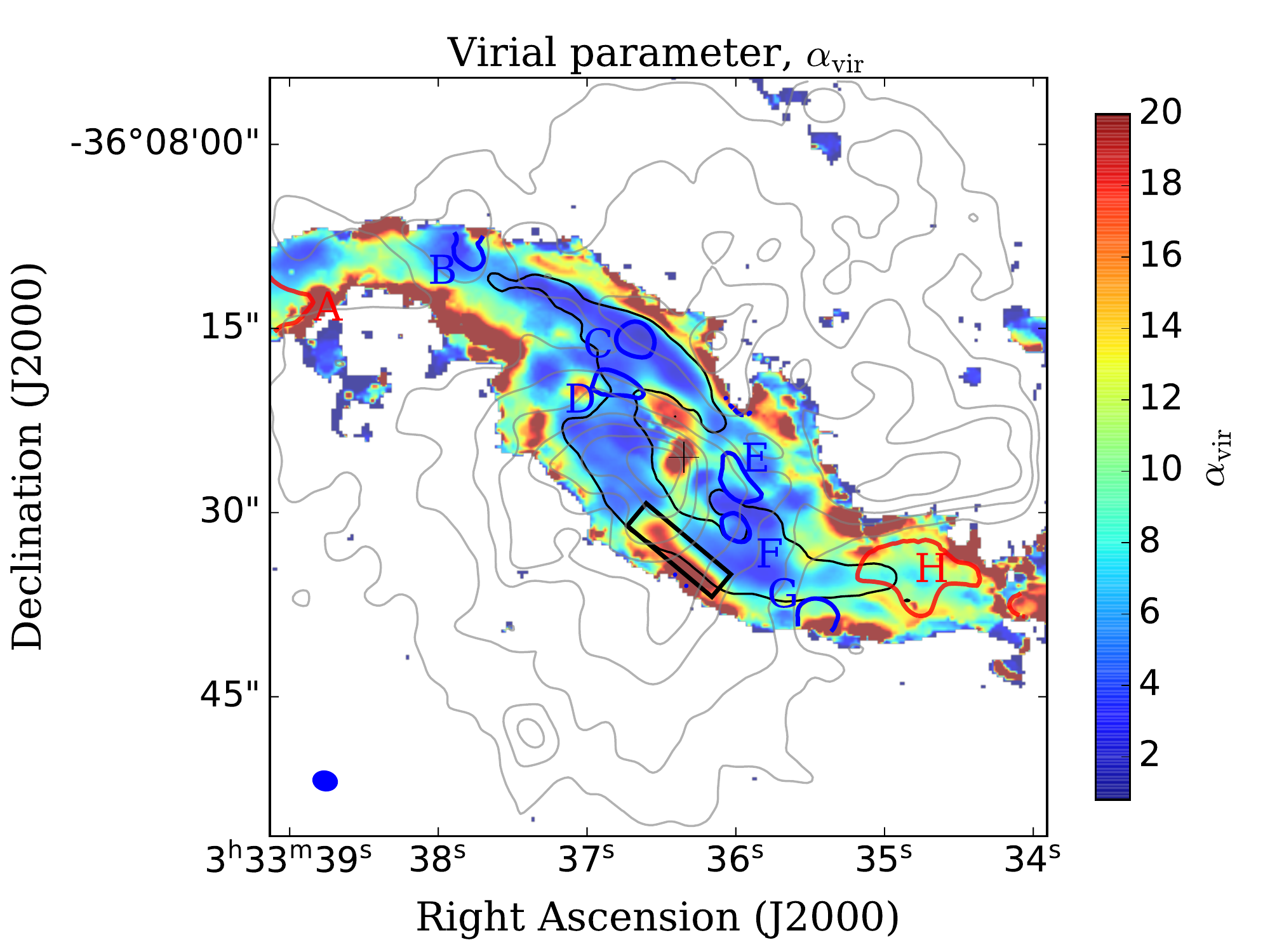}
\caption{$Left-upper$: The SFE map contoured by the $\oiii\lambda5007$ fluxes (gray lines). We mask the central AGN region ($r < 2.4\arcsec$). The black lines represent the dense gas regions with ${\rm log}(\Sigma_{\rm H_2}/\msun \ \rm pc^{-2}) > 2.4$, denoting the shape of dust lanes. The blue and red lines represent some high ($\rm log(SFE/yr^{-1}) > -8.3$) and low ($\rm log(SFE/yr^{-1}) < -11.0$) SFE regions, respectively, which are marked with the letters from A to H. The black box represents the suppressed star-forming region. The typical uncertainty of $\rm log(SFE/yr^{-1})$ is about 0.13. $Right-upper$: The dust attenuation distribution contoured with CO(1-0) fluxes. $Left-bottom$: The moment 2 (velocity dispersion) map of CO(1--0) contoured by the $\oiii\lambda5007$ fluxes (gray lines). $Right-bottom$: The virial parameter $\alpha_{\rm vir}$ distribution of molecular gas. Other lines and symbols are same as the Fig. \ref{fig:SFE}. If we adopt the uncertainty of velocity dispersion as 5 km/s, the typical uncertainty of $\alpha_{\rm vir}$ is less than 0.3.}
\label{fig:SFE}
\end{figure*}

In the strongly barred spiral galaxy NGC 1365, there is an obvious nuclear ring revealed by CO(1-0) (Fig. \ref{fig:alma_muse_data}), CO(2-1) \citep{Sakamoto2007} and CO(3-2) \citep{Combes2018a}, with a radius about 9$^{\arcsec}$ = 770 pc. The position angle of the bar is about 92$^{\circ}$ \citep{Combes2018a}, and the northeastern side is the near side. The location of peak flux in the molecular gas map is consistent with the dust lanes seen in the optical images. The inner Lindblad resonance (ILR) of the bar corresponds to the nuclear ring \citep[e.g., ][]{Sakamoto2007}, where active star formation is taking place.

In Section \ref{subsec:SF}, we present the K-S relation at a resolution of 180 pc, and find a number of enhanced or suppressed star formation regions. In this Section, we smooth the original $\Sigma_{\rm SFR}$ map to match that of the CO(1–0) resolution and investigate the SFE distribution, in order to search for these regions with extraordinary star formation activities. In Fig. \ref{fig:SFE}, we show the SFE map overlaid by the $\oiii\lambda5007$ flux. The dense gas with ${\rm log}(\Sigma_{\rm H_2}/\msun \ \rm pc^{-2}) > 2.4$ are shown by black contours, which roughly depicts the morphology of the two dust lanes. The regions with enhanced star formation ($\rm log(SFE / yr^{-1}) > -8.3$) are accompanied with blue solid lines and marked with letters from B to G. The regions with suppressed star formation ($\rm log(SFE / yr^{-1}) < -11.0$) are represented by the red solid lines and marked A and H. 

Although located in the outer parts of the star fomation ring, regions A and H are found to possess high density molecular gas (${\rm log}(\Sigma_{\rm H_2}/\msun \ \rm pc^{-2}) > 2$) but weak $\ha$ emission. As seen from the dust attenuation distribution (Fig. \ref{fig:SFE}), regions A and H show lower dust attenuation than in B-G regions, which, in contrast, are located where local peaks of $\ha$ flux are found but gas surface density is relatively low. As a result, in the outer regions ($r > 10\arcsec$), a prominent offset between the $\ha$ arms and CO arms is seen, which can be interpreted by the delay of star formation after the compression of molecular gas in spiral arms \citep{Egusa2004, Egusa2009}. The decorrelation between the star formation and the molecular gas leads to the SFE difference between the front/leading side and the back/trailing side of spiral rotation, and even probably result in the breaking down of the resolved K-S law at such a high spatial resolution. 

The regions C, D, E, and F, situating in the inner star-forming ring, show the highest SFE values. However, the gas density within these four regions is of vast difference. The regions C and D are nearly invisible in the optical image, because of the enormous dust extinction attenuation ($A_{\rm v} > 4$). Since the dust/gas column density (${\rm log}(\Sigma_{\rm H_2}/\msun \ \rm pc^{-2}) \sim 2.7$) is extremely high in these two regions, we should note that the extinction correction from the Balmer decrement may not be accurate or even inapplicable \citep{Liu2013}. We defer the comparison of SFR derived from Balmer emission lines and free-free emission of radio continuum to a future work (Gao et al. in prep). In contrast, the regions E and F harbor lower gas surface densities (${\rm log}(\Sigma_{\rm H_2}/\msun \ \rm pc^{-2}) \sim 2.1$), about 25$\%$ of those in C and D. Previously, these four intense star formation regions have been detected in the radio (2, 3, 6, 20 cm) and mid-infrared (8.9 – 12.9 $\mu$m) bands \citep{Sandqvist1982,Sandqvist1995,Galliano2005,Sakamoto2007,Galliano2008,Galliano2012}, which will be discussed later in Section \ref{subsubsec:feedback}.

Furthermore, we note that the SFE in the southwestern (SW) dust lane is lower than that in the northeastern lane by a factor of 5 ($\sim$0.7 dex). In the $left - bottom$ panel of Fig. \ref{fig:SFE}, we plot the velocity dispersion map of CO(1--0) overlaid by the contour of the $\oiii\lambda5007$ flux. We find the molecular gas velocity dispersion in the SW dust lane to be $\sim$50 -- 60 km/s, higher than that in the NE lane. 

The virial parameter, $\alpha_{\rm vir} = M_{\rm vir}/M$, is routinely employed to gauge whether or not a molecular gas cloud fragment is stable against collapse \citep[e.g.,][]{Krumholz2005,Kauffmann2013,Sun2018a}. If $\alpha_{\rm vir} \leq 2$, the cloud fragments are supercritical, unstable and tend to collapse, while $\alpha_{\rm vir} > 2$ suggests that the gas motion alone may prevent cloud fragments from collapsing. Following the method in \cite{Sun2018a}, we calculate $\alpha_{\rm vir}$ at the scale of the beam size through the following equation: 
\begin{equation}
\begin{split}
\alpha_{\rm vir} = \frac{5 \sigma^2 r_{\rm beam}}{f G M} = \frac{5 \rm {ln2}}{\pi f G} (\frac{\sigma}{\rm km/s})^2 \times \\
(\frac{\Sigma}{M_{\odot} \ \rm pc^{-2}})^{-1} (\frac{r_{\rm beam}}{\rm pc})^{-1}. 
\end{split}
\end{equation}
where the value of the factor $f$ is adopted to be 10/9 with a density profile of $\rho(r) \propto r^{-1}$ \citep{Sun2018a}, the gravitational constant $G$ is $4.3 \times 10^{-3} \ {\rm pc} \ \msun^{-1} \ (\rm km/s)^2$, and $r_{\rm beam}$ is the radius of the synthesized beam. We note that the determination of $\alpha_{\rm vir}$ is affected by the adopted value of $X_{\rm CO}$. In Chapter \ref{subsec:almadata}, we adopt a $X_{\rm CO}$ value lower than that of the Milky Way by a factor of 4, but we emphasize that we focus on the comparison of $\alpha_{\rm vir}$ between different sub-galactic regions.

In the $right-bottom$ panel of Fig. \ref{fig:SFE}, we plot the virial parameter distribution of molecular gas at the spatial scale of the beam size. Among these regions, ``C'' shows the smallest $\alpha_{\rm vir}$, in line with with the finding of the highest SFE therein. In addition, $\alpha_{\rm vir}$ is remarkably larger in regions A and H than in B, C, D, E and F, suggestive of low chance for molecular gas to collapse.  Furthermore, significantly high value of $\alpha_{\rm vir}$ is found in the regions on the SW side of the dust lane showing high velocity dispersion, where molecular gas is probably disturbed and heated by the $\oiii\lambda5007$-emitting outflows driven by the central AGN or starburst activities. This result is in line with negative feedback effects from outflows, even in relatively dense gas. In addition, the $\alpha_{\rm vir}$ value in the gas fragments located on the edge of the molecular gas disk is evidently high, which is possibly linked to heating processes, and further discussion on the gas kinematics there is deferred to Section \ref{subsec:kine}.

\subsection{Gas kinematics}
\label{subsec:kine}

\begin{figure*}[ht]
\center
 \includegraphics[width=0.48\textwidth]{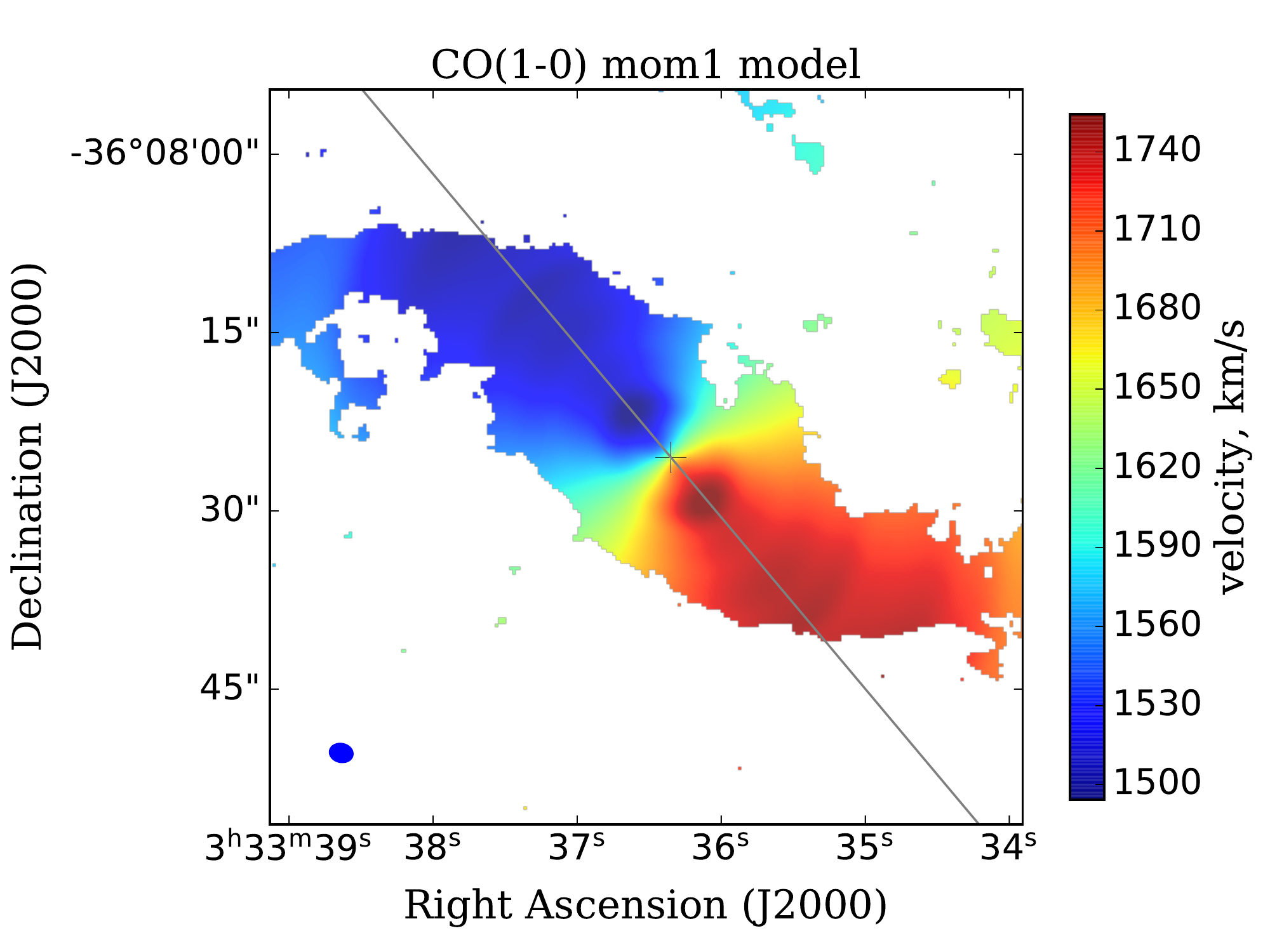}
 \includegraphics[width=0.48\textwidth]{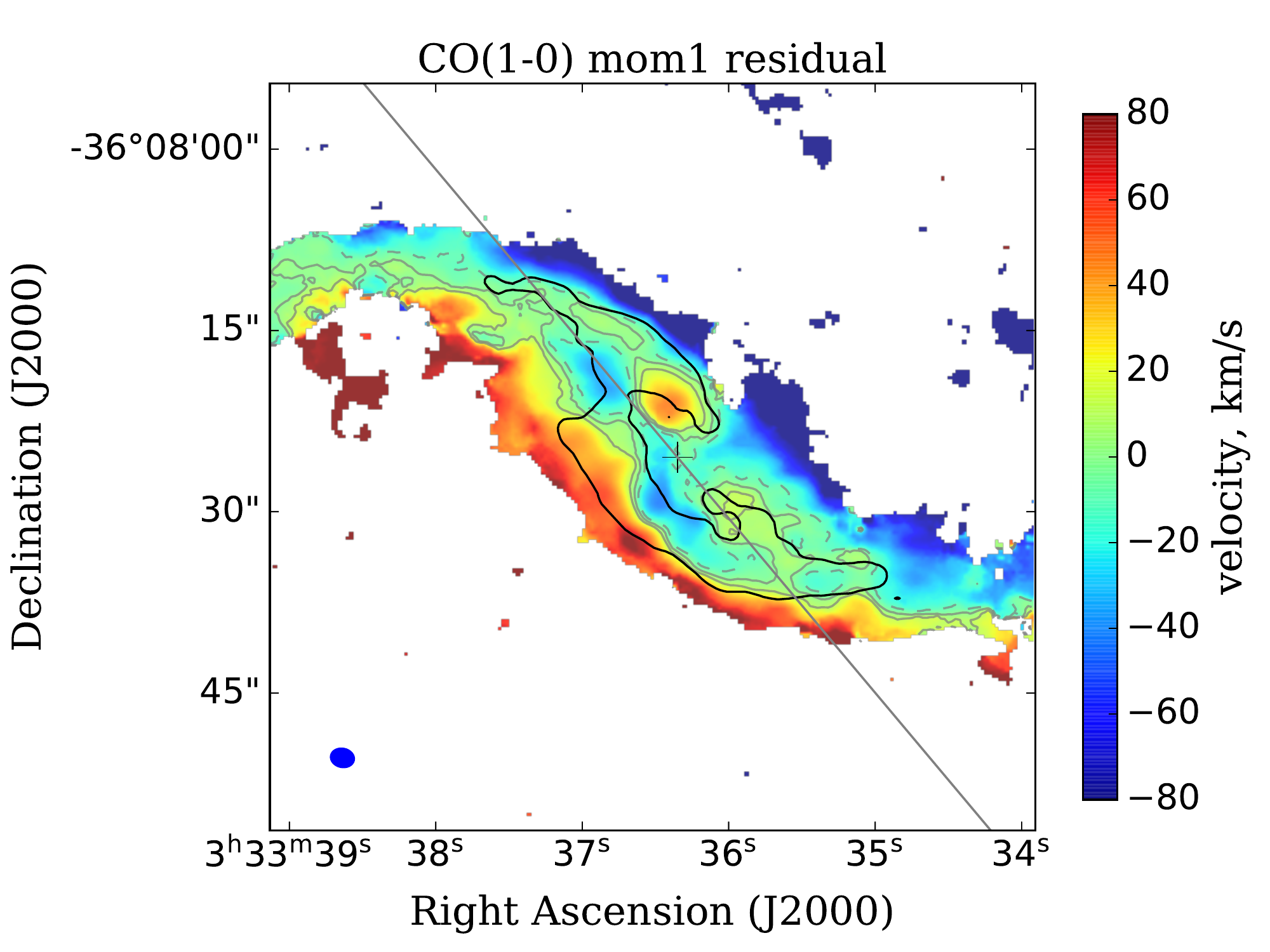}

 \includegraphics[width=0.48\textwidth]{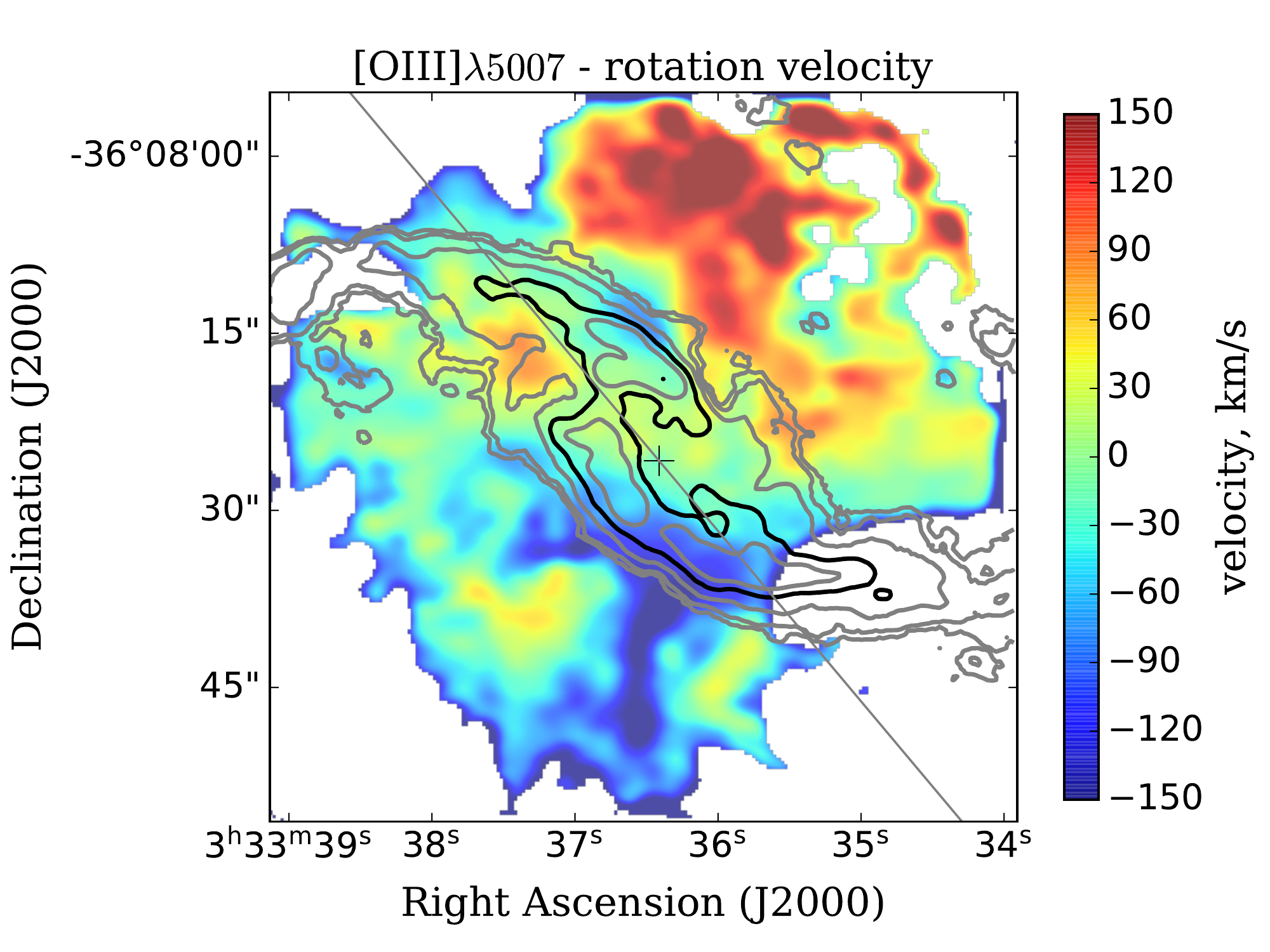}
 \includegraphics[width=0.48\textwidth]{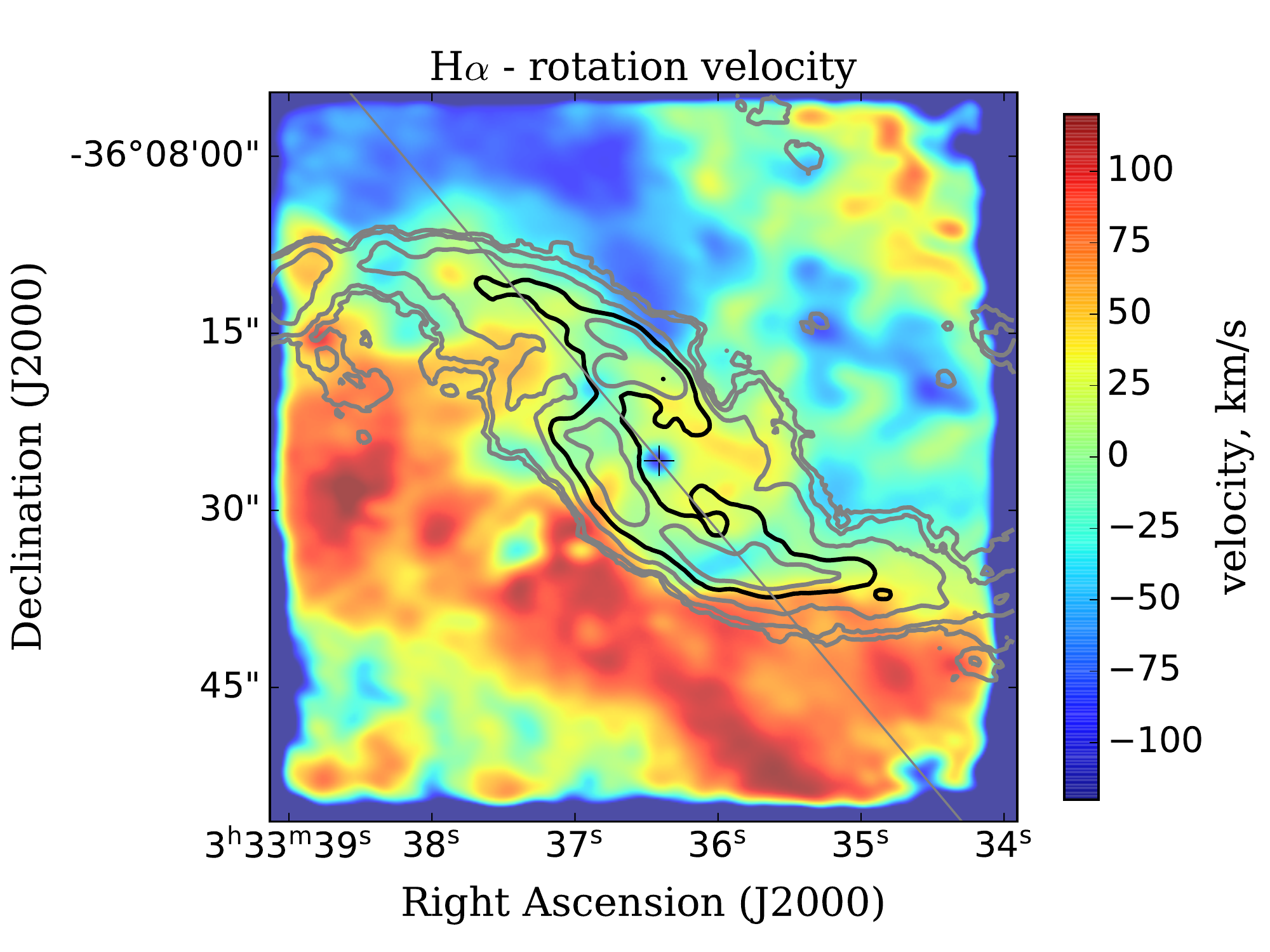}
\caption{$Upper-left$: The rotation model velocity distribution for CO(1--0), computed with code $^{3D}$BAROLO \citep{DiTeodoro2015a}, adopting the systemic velocity of 1618 km/s, PA of 220$^\circ$ and inclination of 40$^\circ$. $Upper-Right$: The velocity residual (observed velocity -- model velocity) map for CO(1--0). The velocity residuals are contoured by the gray lines, separating with [-10, 0, 10] km/s. The negative residuals are shown as dashed lines, while positive values as solid lines. $Bottom$: $\oiii\lambda5007$ ($left$) and $\ha$ ($right$) velocity map subtracted the stellar rotation velocity. They are similar to the Fig. 6 in \cite{Venturi2018a}. The gray contours represent the CO(1--0) emission, which levels are at $(1/2)^n$ ($n = 1, 3, 4, 7$) times of the largest flux. The black lines represent the shape of dust lanes, which is same as the Fig. \ref{fig:SFE}. The gray solid straight line indicates the major axis at PA = 220$^\circ$ and the central AGN is marked with a black cross.}
\label{fig:kine}
\end{figure*}

\begin{figure}
\center
\includegraphics[width=0.45\textwidth]{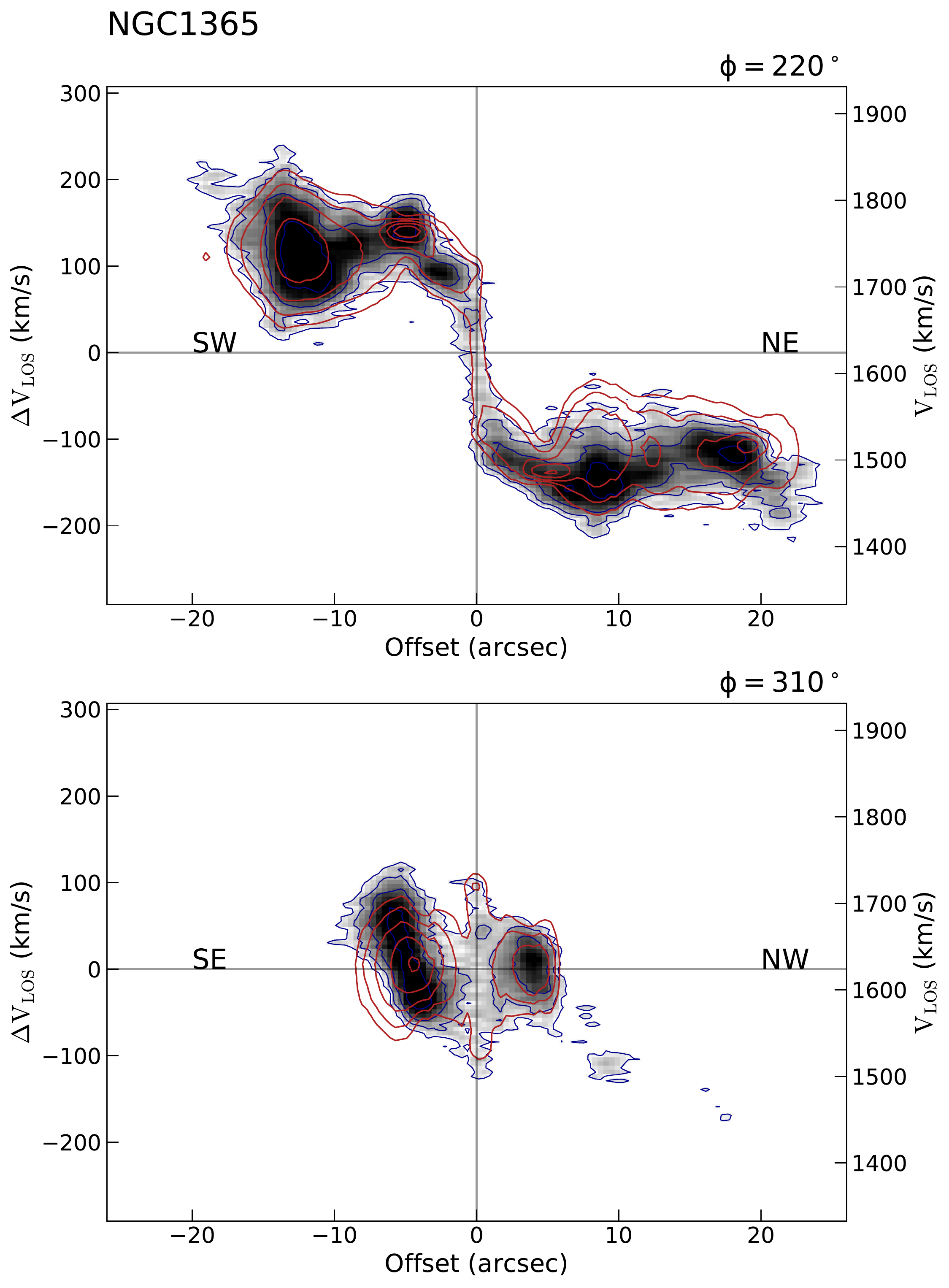}
\caption{The position-velocity diagrams (PVDs) for CO(1--0) emission line along the major axis ($upper$) at PA = 220$^\circ$ and minor axis ($bottom$) at 310$^\circ$. The blue contours represent the observed velocities, which levels are at $2^n\sigma$, from 2$\sigma$ to 64$\sigma$. The best fit models from $^{3D}$BAROLO are shown as red contours. The basic directions are also shown as SW, NE, SE and NW in the plots.}
\label{fig:pvd}
\end{figure}

Previous works have delineated the motion of gas in the nuclear region of NGC 1365 \citep{Lindblad1999, Sakamoto2007, Elmegreen2009}. The molecular gas in the dust lanes flows into the inner region, leading to the formation of massive star clusters and the high gas accretion rate. The high resolution of CO(1--0), $\oiii\lambda5007$ and $\ha$ data now facilitates a scrutinization of the kinematics of the molecular and ionized gas. 

In Fig. \ref{fig:alma_muse_data}, we show the velocity distributions of CO(1--0), $\oiii\lambda5007$ and $\ha$ emission. The velocity map in the right panels therein indicates a non-circular motion of molecular gas, and shows complex motion features around the dust lanes and the star-forming ring. With respect to the systemic velocity (1618 km/s), blueshift in NE and redshift in SW are seen. To model the rotation of molecular gas, we utilize the $^{3D}$BAROLO code \citep{DiTeodoro2015a} to perform a 3D-fitting on the CO(1--0) emission line data cube. The position of the central AGN that we adopt is $\alpha = 03^{\rm h}33^{\rm m}36.35^{\rm s}$, $\delta=-36^{\circ}08{\arcmin}25.8\arcsec$, and the step size of radii ($\Delta r$) is set to be 1$\arcsec$. The systemic velocity, position angle (PA), and inclination are set to be 1618 km/s, 220$^\circ$, and 40$^\circ$, following the parameter setting in \cite{Sakamoto2007}. The $^{3D}$BAROLO code fits a pure circular rotation model to the data cube, so that non-circular motions (e.g., radial motion due to spiral arm and bar dynamics, inflows and outflows) manifest themselves as residuals. The resultant circular rotation model and the residuals are shown in Fig. \ref{fig:kine} ($upper-left$ and $upper-right$ panels, respectively). The residual map shows redshift in the NE dust lane (black solid contour), though blueshift in the SW lane is not evident. As the NW side is the near side of the galaxy \citep{Elmegreen2009}, this indicates streaming inflow motions of molecular gas along the NE dust lane, with a projected velocity of about 10-15 km/s. Interestingly, blueshifted and redshifted strip components outside the dust lanes are readily observed on the NW and SE sides, respectively, which reach up to approximately $\pm$100 km/s in the projected velocity, suggestive of noncircular motion.

Using the velocities derived from emission lines and stellar continuum in Section \ref{subsec:musedata}, we determine the $\oiii\lambda5007$ and $\ha$ velocities relative to the stellar rotation. The stellar rotation is similar to Figure 6 (panel $a$) in \cite{Venturi2018a}. The velocity residual maps derived from the $\oiii\lambda5007$ and $\ha$ emission lines are shown in the $bottom$ panels of Fig. \ref{fig:kine}, which are in consistence with Figure 6 in \cite{Venturi2018a}. In this figure, we see biconical morphology the $\oiii$ outflow, with velocities negative (blueshifted) in the SE, and positive (redshifted) in the NW. Similar receding motion is found on the NW side of the $\ha$ velocity map, though the approaching cone in the SE is unobvious. The $\ha$ velocity with respect to the stellar rotation is dominated by blueshift in the north and redshifted in the south.

We also plot the position-velocity diagrams (PVDs) for CO(1--0) emission in Fig. \ref{fig:pvd}, which are obtained along the major axis at PA = 220$^\circ$ and minor axis at 310$^\circ$. These PVDs help reveal the azimuthal and radial streaming motions along the designated directions \citep{Aalto1999}. The best fit model derived from $^{3D}$BAROLO is shown as red contours. We find that the observed rotation along the major axis has the following piece-wise representation: (1) rigid-body rotation for $r < 2\arcsec$, (2) flat rotation curves for $2\arcsec < r < 20\arcsec$, and (3) additional velocity components. In the minor axis direction, the observed velocity shows a zigzag shape, indicating a velocity gradient along the minor axis. Sole circular motion cannot completely reproduce the observed locus in the $r \sim \pm10\arcsec$ regions, where signatures of outflows in the disk plane likely exist.

\begin{figure}
\center
\includegraphics[width=0.45\textwidth]{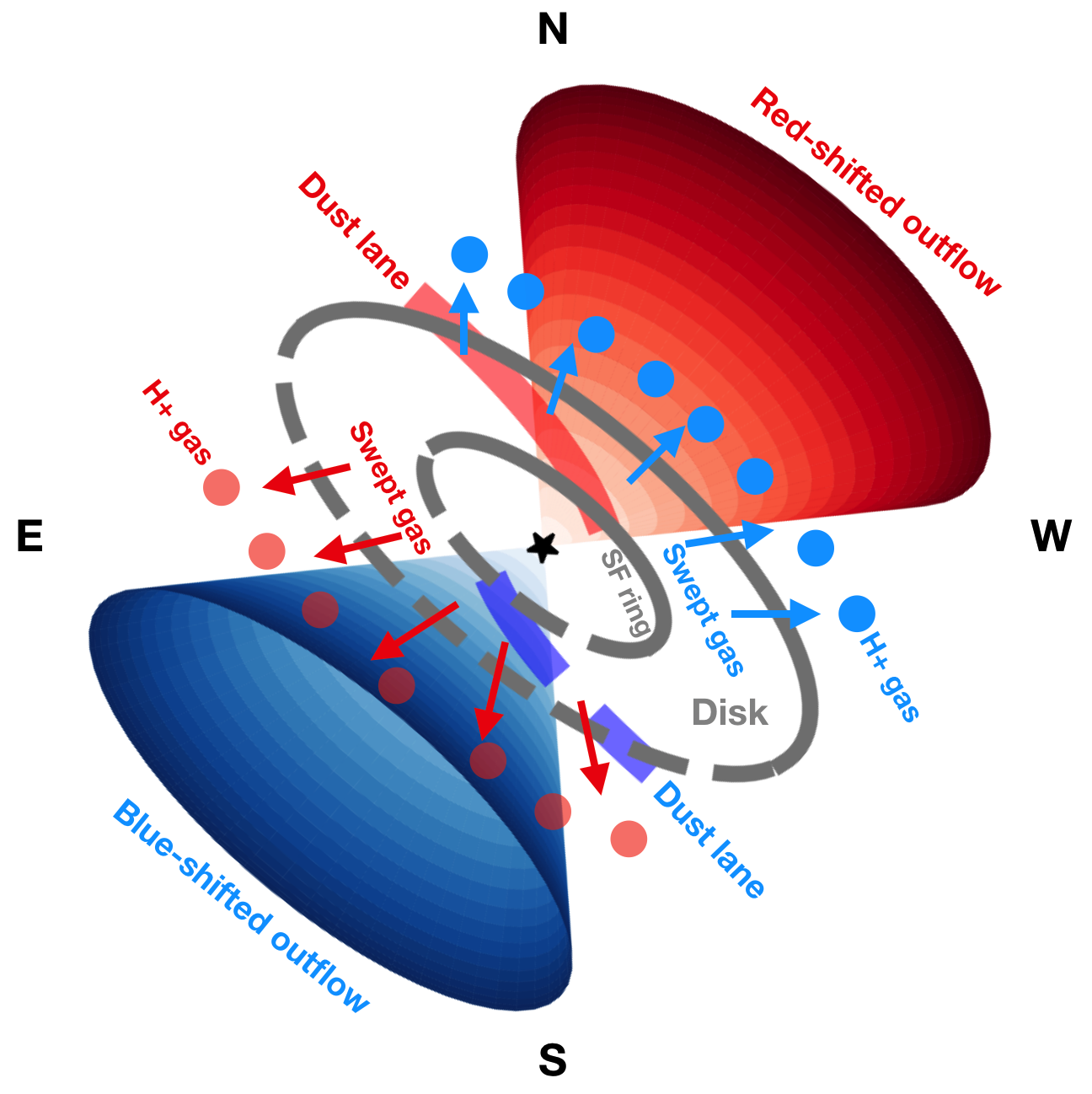}
\caption{Schematic model about kinematics for molecular and ionized gas in the nuclear region of NGC 1365. The inner gray ellipse means the star-forming ring, the region between two gray ellipses represent the molecular disk. The ellipses in front of the biconical outflow are shown as solid, while as a dashed line for the opposite parts. The red (receding or redshifted component) and blue (approaching or blueshifted component) biconical shape represent the outflows of ionized gas triggered by the central AGN and starburst. The inflowing gas in the dust lane, in front of the redshifted outflow, is colored as red, while as blue for another dust lane in behind of the blueshifted outflow. The swept molecular gas in the disk, driven by the outflows, are shown as red (receding) and blue (approaching) arrows. The ionized H$^+$ gas near the disk is shown as blue (approaching) and red (receding) circles. The black star represents the central AGN. }
\label{fig:model}
\end{figure}

The observed velocity maps of molecular and ionized gas are shown in Fig. \ref{fig:alma_muse_data} and Fig. \ref{fig:kine}. Here we provide a schematic view about the kinematics in the nuclear region of NGC 1365, shown in Fig. \ref{fig:model}. The $\oiii$-emitting outflow likely driven by the central AGN and/or starburst is receding in the NW side and approaching on the SE side, and its half-opening angle is as wide as about 50$^\circ$ \citep{Sandqvist1995, Venturi2018a}. The star-forming ring is depicted by the inner gray ellipse, and the molecular disk is located in the region between the two gray ellipses. Dense molecular gas is inflowing to the central AGN along the dust lanes, and is color-coded with red (NW side) and blue (SE side) lines. Along the line of sight, the dust lane on the NW side is in front of the $\oiii$ outflow, while the SE lane is behind the outflow. In Fig. \ref{fig:kine}, the low-density molecular gas in the disk is receding (approaching) on the SE (NW) side, which is shown as red (blue) arrows in Fig. \ref{fig:model}. The low density molecular gas on the surface of the disk may have been swept out by the outflows, and such an interpretation is compatible with their higher virial parameters in Fig. \ref{fig:SFE}. The $\ha$ kinematics in Fig. \ref{fig:kine} is predominantly two thick lanes aligned with the molecular gas whose overall motion is similar (approaching in the NW and receding in the SE ).  The diffuse ionized H$^+$ gas is marked with red and blue circles in Fig. \ref{fig:model}. They are probably created by the photo-ionization of molecular H$_2$ gas and seemingly located between the molecular gas disk and $\oiii$ outflows.

\section{Discussion}
\label{sec:dis}

\subsection{Star formation}
\label{subsec:sf_lane}

The high-resolution data of CO(1--0) and $\ha$ obtained by ALMA and VLT/MUSE allow for exploring the spatially resolved star formation activities in the nuclear region of NGC 1365. In this section, we will discuss the Kennicutt-Schmidt and main-sequence relations at a resolution of 180 pc, and the feedback effects on star formation from the outflows.

\subsubsection{Star formation relations at 180 pc}
\label{subsubsec:relations}

In Section \ref{subsec:SF}, we investigate the K-S relations at a resolution of 180 pc. Because of the higher resolution in our data, the larger slopes are consistent with the findings in \cite{Liu2011}. However, if smoothing the CO and $\ha$ data at a larger scale of 360 pc, we found the correlation coefficients are 0.66 (Pearson) and 0.76 (Spearman), with slopes of 1.64$\pm$0.23 ($\Sigma_{\rm H_2} > 1 \ M_{\odot} \ \rm pc^2$) and 3.0$\pm$0.4 ($\Sigma_{\rm H_2} > 10 \ M_{\odot} \ \rm pc^2$). This super-linear slope ($N \ga 2$) in our work might indicate the transition between inefficient/normal star formation and starburst sequences \citep{Kennicutt2012}. \cite{Onodera2010} studied the K-S law in M33 with different resolutions (80, 240, 500 pc and 1 Kpc) when the $\Sigma_{\rm H_2}$ ranging from 10 to 40 $\msun \rm \ pc^{-2}$, and found the relation becomes invalid and breakdown at the highest resolution of 80 pc. We also detect the breaking down of K-S relation, because the SFR density is nearly uncorrelated with the gas density at the low-density regime.  One reason is that the large fraction of gas at a low molecular gas density regime is atomic $\hi$ gas, because of the low converting efficiency from atomic to molecular phase. The local SFR density is in part uncorrelated with $\hi$ density \citep[e.g., ][]{Liu2011,Kennicutt2012}. Another reason is that the depletion of molecular gas in some inner regions is more efficient, caused by the outflows, starburst, or stellar feedback \citep{Kruijssen2019May}.  Besides, since the delay between compression of molecular gas in spiral arms and the star formation, the decorrelation between the emission peak regions of CO(1--0) and $\ha$ is identified \citep{Egusa2004, Egusa2009}. This might lead to the SFE difference between the front/leading and back/trailing sides of spiral rotation, and even probably contribute the breaking down in the resolved K-S law at such a high spatial resolution. 

In panel $e$ of Fig. \ref{fig:ks_msr}, we find the slope of resolved main sequence relation in the nuclear region of NGC 1365 is steeper than the normal star-forming galaxies and two other spiral galaxies. This difference might be caused by a few reasons. One is that the star formation in nuclear region is enhanced by the strong bar-driven gas dynamics. The stellar bar is suggested to drain the molecular gas in spiral region toward its galactic center, thus trigger the gravitational collapse to form gas clumps and lead to intense star formation \citep[e.g.,][]{Kormendy2004,Chown2019,Lin2020}.
Another is that the coverage of stellar mass density ($7.7 < {\rm log}(\Sigma_*/{M_\odot \ \rm Kpc^{-2}}) < 9.4$) is much narrower than in other studies, which would bring some bias on the relation. The extinction correction is not accurate for dusty regions (e.g. dust lanes), the determination of stellar mass from the SSP fitting might be underestimated at higher densities. Furthermore, the spatial resolution in our work is much higher than in other studies. \cite{Enia2020} derived the resolved main sequence relations for all pixels in eight nearby spiral galaxies at scales of 280 -- 750 pc, and found the slopes are from 0.54 to 1.1, which are smaller than our work. These results indicate that the accumulation speed of stellar mass in the inner region of NGC 1365, such as the star-forming ring, is much faster than in the outer molecular gas spiral arms. 

\subsubsection{Enhanced and suppressed star formation}
\label{subsubsec:feedback}

In Fig. \ref{fig:SFE}, the inner regions C, D, E, and F show the highest SFEs while harboring different gas densities. Regions C and D are nearly invisible in optical bands because they locate at the connection points between the star-forming ring and dust lane. \cite{Galliano2008, Galliano2012} observed three regions (C, D, and another region located at the western with higher dust attenuation value), using instruments of VISIR and SINFONI on VLT in mid-infrared bands.  They regarded these two regions as super young (6 -- 8 Myr) massive ($\sim 10^7 \msun$) star clusters, still embedded by the surrounding dust. \cite{Elmegreen2009} reviewed the environment of these star clusters in NGC 1365. They inferred these clusters are formed at a position where an inter bar filament impacts the dust lane, triggered by the higher pressure and spontaneous gravitational instabilities in the dust lane. The regions E and F have been proven as the mixtures of few optical super star clusters by \cite{Kristen1997}, harboring diffuse molecular gas and two brightest star clusters. \cite{Kristen1997} found the brightest one is 300 times brighter than the luminous globular star clusters in our Milky Way.  \cite{Emsellem2014} provided a detailed simulation about the star formation and gas fueling in the nuclear region of barred galaxies. Their simulation shows that the star clusters would form efficiently at the very edge of the bar and concentrate at the two ends of the bar. Because the end regions of the bar are evolved with a high density of molecular gas while with low shear, the molecular gas is expected to collapse and form the star clusters. The large fraction of the molecular gas in regions E and F have probably been consumed by the star formation or disrupted by stellar feedback and AGN outflows.

In Fig. \ref{fig:SFE},  we notice that the star formation activities in the NE dust lane are much intenser than in the SW one. The denser molecular gas in the SW dust lane is partly heated/interacted by the ionized outflows from the central AGN and starburst regions E and F. Furthermore, \cite{Sandqvist1995} and \cite{Wang2009} have reported the existence of jet in radio and X-ray bands, which features are visually extended to the SW dust lane, possibly leading to the larger velocity dispersion and virial parameters.
This finding supports the negative feedback of outflows \citep{Fabian1994}, because the energy can prevent the molecular gas, even in the denser clouds, from cooling and lead to the suppression of star formation. This result is different from the positive effects of outflows on denser molecular gas clouds \citep[e.g., ][]{Silk2013, Maiolino2017Mar, Shin2019}, which suggest that the outflows will compress the molecular gas and then trigger the star formation activities. In the future, we will check the existence of negative feedback of outflows on denser molecular gas fragments in other galaxies. 

\subsection{Non-circular gas motion} 
\label{subsec:sweep_gas}

As seen in Fig. \ref{fig:kine} and Fig. \ref{fig:model}, we infer that the low-density molecular gas and diffuse ionized gas in the surface of the disk are swept by the outflows of AGN. These motions have also been reported in previous observations \citep[e.g., ][]{GarciaBurillo2014,GarciaBurillo2019,Morganti2015,Salak2016,AlonsoHerrero2019,Shin2019}.  In barred starburst galaxy NGC 1808, \cite{Salak2016} detected these blueshifted and redshifted components of molecular gas in the residual velocity map, which are regarded as the combination of bar dynamics and outflows. \cite{Shin2019} reported that the outflows from AGN in NGC 5728 could sweep out the inflowing gas along the spiral arms, then gave a negative feedback scenario. \cite{GarciaBurillo2019} and \cite{AlonsoHerrero2019} also detected the line-of-sight velocity residuals about 100-200 km/s in NGC 1068 and NGC 3227, respectively. They interpreted the radial motions in the galactic plane as nuclear molecular outflows entrained by the AGN wind. Using simulation, \cite{Emsellem2014} found the stellar feedback from starburst in molecular clumps would expel the gas outside the disk plane to hundreds of parsec. This feedback will help remove the angular momentum from the disk and allow gas to move closer to the inner AGN. Except for the sweeping motion, the inflowing gas along the dust lanes is detected in the CO(1--0) velocity residual map (Fig. \ref{fig:kine}), which is positive in NE dust lane while is not evident in SW one, because these components are smaller in terms of velocity residuals compared to the outflowing components.

Another possibility is that the complex bar dynamics (at least partially) contribute to these non-circular motions \citep{Koda2006}. However, to derive an accurate picture of the kinematics of stellar, ionized gas, and molecular gas, we need to perform a simulation about the rotation motions of spiral arms, bar, and star-forming ring \citep{Sakamoto1999, Koda2006, Emsellem2014, Li2015a}, which will be studied in our future work.

\subsection{Mass outflow rate and powering source}

\begin{deluxetable}{llr}
\tablecaption{Parameters of outflow \label{tab:para_outflow}}
\tablehead{
\colhead{Parameters} & \colhead{Values} & \colhead{Reference}  }
\startdata
$M_{\rm Mol}$ (total)  & $\sim 1.91 \times 10^9 \ M_{\odot}$ & This work \\
$M_{\rm Mol}$ (not outflow)  & $\sim 1.78 \times 10^9 \ M_{\odot}$ & This work \\
$M_{\rm Mol}$ (outflow) & $\sim 1.37 \times 10^8 \ M_{\odot}$ & This work \\
$R_{\rm out}$  & $\sim 598$ pc  					   & This work \\
$V_{\rm out}$  & $73^{+35}_{-17} \ \rm km/s$  	   & This work \\
${\rm d}M/{\rm d}t$  & $35^{+16}_{-8} \ M_{\odot} \rm \ yr^{-1}$ & This work \\
SFR (total) 		 & $\sim$ 17.0 $M_{\odot} \rm \ yr^{-1}$        & \cite{Combes2018a} \\
SFR (central) 		 & $\sim$ 4.6 $M_{\odot} \rm \ yr^{-1}$         & This work \\
${\rm d}P_{\rm out}/{\rm d}t$ & $2.1^{+2.4}_{-0.9} \times 10^{34} \ \rm g \ cm \ s^{-2}$ & This work \\
$L_{\rm bol}/c$ & $\sim 6.7 \times 10^{32} \ \rm g \ cm \ s^{-2}$   & This work \\
$L_{\rm kin}$   & $1.0^{+2.2}_{-0.5} \times 10^{41} \ \rm erg \ s^{-1}$ & This work \\
$L_{\rm bol}$   & $\sim 2 \times 10^{43} \ {\rm erg \ s^{-1}}$      & \cite{Venturi2018a} 
\enddata
\tablecomments{The total molecular gas mass in central 5 Kpc region is represented by $M_{\rm Mol}$ (total). The $M_{\rm Mol}$ (not outflow) means the molecular gas mass with  $-50 \ {\rm km/s} < V - V_{\rm sys} < 50 \ {\rm km/s}$. The adopted conversion factor $X_{\rm CO}$ is $\rm 0.5 \times 10^{20} cm^{-2} (K \ km \ s^{-1})^{-1}$, which is same as in Section \ref{subsec:almadata}. SFR (total) in NGC 1365 is derived from the infrared luminosities.  SFR (central) means the total SFR in central 5 Kpc region, derived from the attenuation-corrected $\ha$ luminosity. }
\end{deluxetable}

With the spatial distribution of molecular gas and its velocity, it is easy to derive the mass of outflowing molecular gas ($M_{\rm Mol}$), projected radial size ($R_{\rm out}$), and projected outflowing velocity ($V_{\rm out}$). After subtracting the contribution of molecular gas with $-50 \ {\rm km/s} < V - V_{\rm sys} < 50 \ {\rm km/s}$, we obtain the $M_{\rm Mol}$, $R_{\rm out}$, and $V_{\rm out}$. Here, we use the conversion factor $X_{\rm CO}$ in starburst nuclei, which is same as in Section \ref{subsec:almadata}. Following the procedure in previous work \cite[Eq. 4, 6, 7; ][]{GarciaBurillo2014} and adopting the angle $\alpha$ between the outflow and the line of sight as 40$^\circ$, we estimate the mass outflow rate (${\rm d}M/{\rm d}t$), the kinetic luminosity ($L_{\rm kin}$) of outflow, and the momentum flux (${\rm d}P_{\rm out}/{\rm d}t$) of outflow. The total SFR of NGC 1365 is adopted from literature and the SFR in the central 5 Kpc region is estimated from attenuation-corrected $\ha$ luminosity. We also obtain the momentum ($L_{\rm bol}/c$) provided by AGN. All of these parameters are listed in Table \ref{tab:para_outflow}.

The $L_{\rm kin}/L_{\rm bol}$ is about $0.5^{+1.1}_{-0.3}\%$, which is much lower than the required fraction of $5\% L_{\rm bol}$ in AGN feedback model to produce an outflow in ISM \citep{DiMatteo2005,King2015}, and similar to fraction of $0.5\%$ in the two-phase feedback model \citep{Hopkins2010}. These results suggest that the energy of AGN is enough to produce such an outflow. The $({\rm d}P_{\rm out}/{\rm d}t)/(L_{\rm bol}/c) \sim 31^{+37}_{-12}$ is nearly consistent with the range of momentum boost factors ($({\rm d}P_{\rm out}/{\rm d}t)/(L_{\rm bol}/c) \sim 10-50$) in the AGN feedback model with energy-conserving outflows predicted by \cite{FaucherGiguere2012}. Accoring to the explanation in \cite{FaucherGiguere2012}, these results indicate that the AGN nuclear winds or hot shocked gas probably be the primary driving mechanism of molecular outflow. Furthermore, the $({\rm d}M/{\rm d}t)/\rm SFR$ are about $2.0^{+1.0}_{-0.4}$ and $7.5^{+3.8}_{-1.7}$ for the global galaxy and the central 5 Kpc region, respectively. The consumption of molecular gas via outflowing is much faster than star formation, suggesting the negative feedback scenario.

\section{Summary}
\label{sec:sum}

In this work, we perform a spatially resolved analysis of molecular gas and ionized gas in the central 5.4 $\times$ 5.4 Kpc region of NGC 1365, using the ALMA band 3 and VLT/MUSE data. We explore the star formation activities and kinematics in dust lanes, circumnuclear ring, and outflow biconical regions. The main conclusions are summarized below. 

\begin{itemize}

\item We find the resolved K-S relation is super-linear at a resolution of 180 pc, with steeper slopes than previous studies based on larger spatial resolution scales. We suggest the large slopes reflect the transition between normal star formation and starburst sequences. The star formation activities in the inner circumnuclear ring are intenser than in outer regions. 

\item The slope of resolved main sequence relation in the nuclear region of NGC 1365 is steeper than in the normal star-forming galaxies and other spiral galaxies at a smaller spatial resolution. This indicates that the accumulation speed of stellar mass in the inner region, such as star-forming ring, is much faster than in the outer spiral arms, suggesting the enhancement of bar dynamics on star formation. 

\item The regions C, D, E, and F in the inner star-forming ring show the highest star formation efficiency. These regions are regarded to harbor massive star clusters and might be caused by the cloud-cloud collisions in the denser molecular gas regime. 

\item The star formation in the SW dust lane is much weaker than the NE one while harboring denser molecular gas, larger velocity dispersion, and virial parameters. The SW dust lane is also superposed at the larger $\oiii\lambda5007$ velocity region. These results suggest the scenario of negative feedback of outflows, because the radiation energy/outflows from the central AGN and starburst can prevent the molecular gas from cooling even in the denser clouds. 

\item After subtracting a circular molecular gas rotation model and the stellar rotation, we find two obvious non-circular motion components of molecular and ionized hydrogen gas, reaching velocity up to 100 km/s. These motions probably indicate the scenario that the outflows from AGN could sweep out the low-density molecular gas and diffuse ionized gas on the surface of the disk.

\item The molecular outflow is probably driven by AGN nuclear winds or hot shocked gas. The consumption of molecular gas via outflowing is faster than star formation, suggesting the negative feedback scenario.
\end{itemize}

\acknowledgments
We thank the referee very much for his/her careful reading and valuable suggestions.
This work is supported by the grant from the National Key R$\&$D Program of China (2016YFA0400702), the National Natural Science Foundation of China (No. 11673020 and No. 11421303), the Fundamental Research Funds for the Central Universities, and the Chinese Space Station Telescope (CSST) Project. 
Y.L.G. gratefully acknowledges support from the China Scholarship Council (No. 201906340095).  
F.E. is supported by JSPS KAKENHI Grant Number 17K14259.  K.M.M. is supported by JSPS KAKENHI Grant Numbers 19J40004 and 19H01931. We thank Drs. Junzhi Wang, Zhiyu Zhang, Yu Gao, Junfeng Wang, Zhenyi Cai, and Qiusheng Gu for the fruitful discussion and advice. This paper makes use of the following ALMA data: ADS/JAO.ALMA$\#$ 2015.1.01135.S, 2017.1.00129.S. ALMA is a partnership of ESO (representing its member states), NSF (USA), and NINS (Japan), together with NRC (Canada) and NSC and ASIAA (Taiwan), in cooperation with the Republic of Chile. The Joint ALMA Observatory is operated by ESO, AUI/NRAO, and NAOJ. The National Radio Astronomy Observatory is a facility of the National Science Foundation operated under cooperative agreement by Associated Universities, Inc.


\facilities{ALMA, VLT/MUSE}
\software{CASA, Python, $^{3D}$BAROLO, MPFIT}

\bibliography{ngc1365}

\begin{thebibliography}{}
\expandafter\ifx\csname natexlab\endcsname\relax\def\natexlab#1{#1}\fi
\providecommand{\url}[1]{\href{#1}{#1}}
\providecommand{\dodoi}[1]{doi:~\href{http://doi.org/#1}{\nolinkurl{#1}}}
\providecommand{\doeprint}[1]{\href{http://ascl.net/#1}{\nolinkurl{http://ascl.net/#1}}}
\providecommand{\doarXiv}[1]{\href{https://arxiv.org/abs/#1}{\nolinkurl{https://arxiv.org/abs/#1}}}

\bibitem[{Aalto {et~al.}(1999)Aalto, H{\"u}ttemeister, Scoville, \&
  Thaddeus}]{Aalto1999}
Aalto, S., H{\"u}ttemeister, S., Scoville, N.~Z., \& Thaddeus, P. 1999, \apj,
  522, 165, \dodoi{10.1086/307610}

\bibitem[{Agostino \& Salim(2019)}]{Agostino2019}
Agostino, C.~J., \& Salim, S. 2019, \apj, 876, 12,
  \dodoi{10.3847/1538-4357/ab1094}

\bibitem[{Alonso-Herrero {et~al.}(2019)Alonso-Herrero, Garc{\'\i}a-Burillo,
  Pereira-Santaella, Davies, Combes, Vestergaard, Raimundo, Bunker,
  D{\'\i}az-Santos, Gandhi, Garc{\'\i}a-Bernete, Hicks, H{\"o}nig, Hunt,
  Imanishi, Izumi, Levenson, Maciejewski, Packham, Ramos~Almeida, Ricci,
  Rigopoulou, Roche, Rosario, Schartmann, Usero, \& Ward}]{AlonsoHerrero2019}
Alonso-Herrero, A., Garc{\'\i}a-Burillo, S., Pereira-Santaella, M., {et~al.}
  2019, \aap, 628, A65, \dodoi{10.1051/0004-6361/201935431}

\bibitem[{Azeez {et~al.}(2016)Azeez, Hwang, Abidin, \& Ibrahim}]{Azeez2016}
Azeez, J.~H., Hwang, C.-Y., Abidin, Z.~Z., \& Ibrahim, Z.~A. 2016, Scientific
  Reports, 6, 26896, \dodoi{10.1038/srep26896}

\bibitem[{Bacon {et~al.}(2016)Bacon, Piqueras, Conseil, Richard, \&
  Shepherd}]{Bacon2016}
Bacon, R., Piqueras, L., Conseil, S., Richard, J., \& Shepherd, M. 2016, MPDAF:
  MUSE Python Data Analysis Framework.
\newblock \doeprint{1611.003}

\bibitem[{Bacon {et~al.}(2010)Bacon, Accardo, Adjali, Anwand, Bauer, Biswas,
  Blaizot, Boudon, Brau-Nogue, Brinchmann, Caillier, Capoani, Carollo, Contini,
  Couderc, Daguis{\'e}, Deiries, Delabre, Dreizler, Dubois, Dupieux, Dupuy,
  Emsellem, Fechner, Fleischmann, Fran{\c{c}}ois, Gallou, Gharsa, Glindemann,
  Gojak, Guiderdoni, Hansali, Hahn, Jarno, Kelz, Koehler, Kosmalski, Laurent,
  Le~Floch, Lilly, Lizon, Loupias, Manescau, Monstein, Nicklas, Olaya, Pares,
  Pasquini, P{\'e}contal-Rousset, Pell{\'o}, Petit, Popow, Reiss, Remillieux,
  Renault, Roth, Rupprecht, Serre, Schaye, Soucail, Steinmetz, Streicher,
  Stuik, Valentin, Vernet, Weilbacher, Wisotzki, \& Yerle}]{Bacon2010}
Bacon, R., Accardo, M., Adjali, L., {et~al.} 2010, Society of Photo-Optical
  Instrumentation Engineers (SPIE) Conference Series, Vol. 7735, The MUSE
  second-generation VLT instrument, 773508, \dodoi{10.1117/12.856027}

\bibitem[{Baldwin {et~al.}(1981)Baldwin, Phillips, \& Terlevich}]{Baldwin1981}
Baldwin, A., Phillips, M.~M., \& Terlevich, R. 1981, \pasp, 93, 817,
  \dodoi{10.1086/130930}

\bibitem[{Bigiel {et~al.}(2008)Bigiel, Leroy, Walter, Brinks, de~Blok, Madore,
  \& Thornley}]{Bigiel2008}
Bigiel, F., Leroy, A., Walter, F., {et~al.} 2008, \aj, 136, 2846,
  \dodoi{10.1088/0004-6256/136/6/2846}

\bibitem[{Bolatto {et~al.}(2013)Bolatto, Wolfire, \& Leroy}]{Bolatto2013}
Bolatto, A.~D., Wolfire, M., \& Leroy, A.~K. 2013, \araa, 51, 207,
  \dodoi{10.1146/annurev-astro-082812-140944}

\bibitem[{Bruzual \& Charlot(2003)}]{Bruzual2003}
Bruzual, G., \& Charlot, S. 2003, \mnras, 344, 1000,
  \dodoi{10.1046/j.1365-8711.2003.06897.x}

\bibitem[{Calzetti {et~al.}(2000)Calzetti, Armus, Bohlin, Kinney, Koornneef, \&
  Storchi-Bergmann}]{Calzetti2000}
Calzetti, D., Armus, L., Bohlin, R.~C., {et~al.} 2000, \apj, 533, 682,
  \dodoi{10.1086/308692}

\bibitem[{Chabrier(2003)}]{Chabrier2003}
Chabrier, G. 2003, \pasp, 115, 763, \dodoi{10.1086/376392}

\bibitem[{Cheung {et~al.}(2016)Cheung, Bundy, Cappellari, Peirani, Rujopakarn,
  Westfall, {et~al.}}]{Cheung2016May}
Cheung, E., Bundy, K., Cappellari, M., {et~al.} 2016, 533, 504,
  \dodoi{10.1038/nature18006}

\bibitem[{Chown {et~al.}(2019)Chown, Li, Athanassoula, Li, Wilson, Lin, Mo,
  Parker, \& Xiao}]{Chown2019}
Chown, R., Li, C., Athanassoula, E., {et~al.} 2019, \mnras, 484, 5192,
  \dodoi{10.1093/mnras/stz349}

\bibitem[{Cid~Fernandes {et~al.}(2005)Cid~Fernandes, Mateus, Sodr{\'e},
  Stasi{\'n}ska, \& Gomes}]{CidFernandes2005}
Cid~Fernandes, R., Mateus, A., Sodr{\'e}, L., Stasi{\'n}ska, G., \& Gomes,
  J.~M. 2005, \mnras, 358, 363, \dodoi{10.1111/j.1365-2966.2005.08752.x}

\bibitem[{Combes {et~al.}(2019)Combes, Garcia-Burillo, Audibert, Hunt, Eckart,
  Aalto, Casasola, Boone, Krips, Viti, Sakamoto, Muller, Dasyra, van~der Werf,
  \& Martin}]{Combes2018a}
Combes, F., Garcia-Burillo, S., Audibert, A., {et~al.} 2019, \aap, 623, A79,
  \dodoi{10.1051/0004-6361/201834560}

\bibitem[{Cresci {et~al.}(2015)Cresci, Marconi, Zibetti, Risaliti, Carniani,
  Mannucci, Gallazzi, Maiolino, Balmaverde, Brusa, Capetti, Cicone, Feruglio,
  Bland-Hawthorn, Nagao, Oliva, Salvato, Sani, Tozzi, Urrutia, \&
  Venturi}]{Cresci2015}
Cresci, G., Marconi, A., Zibetti, S., {et~al.} 2015, \aap, 582, A63,
  \dodoi{10.1051/0004-6361/201526581}

\bibitem[{Davies {et~al.}(2016)Davies, Groves, Kewley, Dopita, Hampton,
  Shastri, Scharw{\"a}chter, Sutherland, Kharb, Bhatt, Jin, Banfield, Zaw,
  James, Juneau, \& Srivastava}]{Davies2016}
Davies, R.~L., Groves, B., Kewley, L.~J., {et~al.} 2016, \mnras, 462, 1616,
  \dodoi{10.1093/mnras/stw1754}

\bibitem[{Di~Matteo {et~al.}(2005)Di~Matteo, Springel, \&
  Hernquist}]{DiMatteo2005}
Di~Matteo, T., Springel, V., \& Hernquist, L. 2005, \nat, 433, 604,
  \dodoi{10.1038/nature03335}

\bibitem[{Di~Teodoro \& Fraternali(2015)}]{DiTeodoro2015a}
Di~Teodoro, E.~M., \& Fraternali, F. 2015, \mnras, 451, 3021,
  \dodoi{10.1093/mnras/stv1213}

\bibitem[{Dobbs(2008)}]{Dobbs2008}
Dobbs, C.~L. 2008, \mnras, 391, 844, \dodoi{10.1111/j.1365-2966.2008.13939.x}

\bibitem[{Dobbs(2014)}]{Dobbs2014}
Dobbs, C.~L. 2014, in IAU Symposium, Vol. 298, Setting the scene for Gaia and
  LAMOST, ed. S.~{Feltzing}, G.~{Zhao}, N.~A. {Walton}, \& P.~{Whitelock},
  221--227, \dodoi{10.1017/S1743921313006406}

\bibitem[{Durr{\'e} \& Mould(2018)}]{Durre2018}
Durr{\'e}, M., \& Mould, J. 2018, \apj, 867, 149,
  \dodoi{10.3847/1538-4357/aae68e}

\bibitem[{Eden {et~al.}(2012)Eden, Moore, Plume, \& Morgan}]{Eden2012}
Eden, D.~J., Moore, T. J.~T., Plume, R., \& Morgan, L.~K. 2012, \mnras, 422,
  3178, \dodoi{10.1111/j.1365-2966.2012.20840.x}

\bibitem[{Eden {et~al.}(2015)Eden, Moore, Urquhart, Elia, Plume, Rigby, \&
  Thompson}]{Eden2015}
Eden, D.~J., Moore, T. J.~T., Urquhart, J.~S., {et~al.} 2015, \mnras, 452, 289,
  \dodoi{10.1093/mnras/stv1323}

\bibitem[{Egusa {et~al.}(2009)Egusa, Kohno, Sofue, Nakanishi, \&
  Komugi}]{Egusa2009}
Egusa, F., Kohno, K., Sofue, Y., Nakanishi, H., \& Komugi, S. 2009, \apj, 697,
  1870, \dodoi{10.1088/0004-637X/697/2/1870}

\bibitem[{Egusa {et~al.}(2004)Egusa, Sofue, \& Nakanishi}]{Egusa2004}
Egusa, F., Sofue, Y., \& Nakanishi, H. 2004, \pasj, 56, L45,
  \dodoi{10.1093/pasj/56.6.L45}

\bibitem[{Elmegreen \& Elmegreen(1983)}]{Elmegreen1983}
Elmegreen, B.~G., \& Elmegreen, D.~M. 1983, \mnras, 203, 31,
  \dodoi{10.1093/mnras/203.1.31}

\bibitem[{Elmegreen \& Elmegreen(2019)}]{Elmegreen2019}
---. 2019, \apjs, 245, 14, \dodoi{10.3847/1538-4365/ab4903}

\bibitem[{Elmegreen {et~al.}(2009)Elmegreen, Galliano, \&
  Alloin}]{Elmegreen2009}
Elmegreen, B.~G., Galliano, E., \& Alloin, D. 2009, \apj, 703, 1297,
  \dodoi{10.1088/0004-637X/703/2/1297}

\bibitem[{Emsellem {et~al.}(2015)Emsellem, Renaud, Bournaud, Elmegreen, Combes,
  \& Gabor}]{Emsellem2014}
Emsellem, E., Renaud, F., Bournaud, F., {et~al.} 2015, \mnras, 446, 2468,
  \dodoi{10.1093/mnras/stu2209}

\bibitem[{Enia {et~al.}(2020)Enia, Rodighiero, Morselli, Casasola, Bianchi,
  Rodriguez-Mu{\~n}oz, Mancini, Renzini, Popesso, Cassata, Negrello, \&
  Franceschini}]{Enia2020}
Enia, A., Rodighiero, G., Morselli, L., {et~al.} 2020, \mnras, 493, 4107,
  \dodoi{10.1093/mnras/staa433}

\bibitem[{Evans {et~al.}(2014)Evans, Heiderman, \& Vutisalchavakul}]{Evans2014}
Evans, Neal~J., I.~I., Heiderman, A., \& Vutisalchavakul, N. 2014, \apj, 782,
  114, \dodoi{10.1088/0004-637X/782/2/114}

\bibitem[{Fabian(1994)}]{Fabian1994}
Fabian, A.~C. 1994, \araa, 32, 277, \dodoi{10.1146/annurev.aa.32.090194.001425}

\bibitem[{Fabian(2012)}]{Fabian2012}
---. 2012, \araa, 50, 455, \dodoi{10.1146/annurev-astro-081811-125521}

\bibitem[{Faucher-Gigu{\`e}re \& Quataert(2012)}]{FaucherGiguere2012}
Faucher-Gigu{\`e}re, C.-A., \& Quataert, E. 2012, \mnras, 425, 605,
  \dodoi{10.1111/j.1365-2966.2012.21512.x}

\bibitem[{Federrath \& Klessen(2012)}]{Federrath2012}
Federrath, C., \& Klessen, R.~S. 2012, \apj, 761, 156,
  \dodoi{10.1088/0004-637X/761/2/156}

\bibitem[{Foyle {et~al.}(2010)Foyle, Rix, Walter, \& Leroy}]{Foyle2010}
Foyle, K., Rix, H.~W., Walter, F., \& Leroy, A.~K. 2010, \apj, 725, 534,
  \dodoi{10.1088/0004-637X/725/1/534}

\bibitem[{Gallagher {et~al.}(2019)Gallagher, Maiolino, Belfiore, Drory, Riffel,
  \& Riffel}]{Gallagher2019}
Gallagher, R., Maiolino, R., Belfiore, F., {et~al.} 2019, \mnras, 485, 3409,
  \dodoi{10.1093/mnras/stz564}

\bibitem[{Galliano {et~al.}(2008)Galliano, Alloin, Pantin, Granato, Delva,
  Silva, Lagage, \& Panuzzo}]{Galliano2008}
Galliano, E., Alloin, D., Pantin, E., {et~al.} 2008, \aap, 492, 3,
  \dodoi{10.1051/0004-6361:20077621}

\bibitem[{Galliano {et~al.}(2005)Galliano, Alloin, Pantin, Lagage, \&
  Marco}]{Galliano2005}
Galliano, E., Alloin, D., Pantin, E., Lagage, P.~O., \& Marco, O. 2005, \aap,
  438, 803, \dodoi{10.1051/0004-6361:20053049}

\bibitem[{Galliano {et~al.}(2012)Galliano, Kissler-Patig, Alloin, \&
  Telles}]{Galliano2012}
Galliano, E., Kissler-Patig, M., Alloin, D., \& Telles, E. 2012, \aap, 545,
  A10, \dodoi{10.1051/0004-6361/201218812}

\bibitem[{Gao {et~al.}(2018)Gao, Bao, Yuan, Kong, Zou, Zhou,
  {et~al.}}]{Gao2018Dec}
Gao, Y., Bao, M., Yuan, Q., {et~al.} 2018, \apj, 869, 15,
  \dodoi{10.3847/1538-4357/aae9ef}

\bibitem[{Gao \& Solomon(2004)}]{Gao2004}
Gao, Y., \& Solomon, P.~M. 2004, \apj, 606, 271, \dodoi{10.1086/382999}

\bibitem[{Garc{\'\i}a-Burillo {et~al.}(2012)Garc{\'\i}a-Burillo, Usero,
  Alonso-Herrero, Graci{\'a}-Carpio, Pereira-Santaella, Colina, Planesas, \&
  Arribas}]{GarciaBurillo2012}
Garc{\'\i}a-Burillo, S., Usero, A., Alonso-Herrero, A., {et~al.} 2012, \aap,
  539, A8, \dodoi{10.1051/0004-6361/201117838}

\bibitem[{Garc{\'\i}a-Burillo {et~al.}(2014)Garc{\'\i}a-Burillo, Combes, Usero,
  Aalto, Krips, Viti, Alonso-Herrero, Hunt, Schinnerer, Baker, Boone, Casasola,
  Colina, Costagliola, Eckart, Fuente, Henkel, Labiano, Mart{\'\i}n,
  M{\'a}rquez, Muller, Planesas, Ramos~Almeida, Spaans, Tacconi, \& van~der
  Werf}]{GarciaBurillo2014}
Garc{\'\i}a-Burillo, S., Combes, F., Usero, A., {et~al.} 2014, \aap, 567, A125,
  \dodoi{10.1051/0004-6361/201423843}

\bibitem[{Garc{\'\i}a-Burillo {et~al.}(2019)Garc{\'\i}a-Burillo, Combes,
  Ramos~Almeida, Usero, Alonso-Herrero, Hunt, Rouan, Aalto, Querejeta, Viti,
  van~der Werf, Vives-Arias, Fuente, Colina, Mart{\'\i}n-Pintado, Henkel,
  Mart{\'\i}n, Krips, Gratadour, Neri, \& Tacconi}]{GarciaBurillo2019}
Garc{\'\i}a-Burillo, S., Combes, F., Ramos~Almeida, C., {et~al.} 2019, \aap,
  632, A61, \dodoi{10.1051/0004-6361/201936606}

\bibitem[{Genzel {et~al.}(2020)Genzel, Price, {\"U}bler, F{\"o}rster~Schreiber,
  Shimizu, Tacconi, Bender, Burkert, Contursi, Coogan, Davies, Davies, Dekel,
  Herrera-Camus, Lee, Lutz, Naab, Neri, Nestor, Renzini, Saglia, Schuster,
  Sternberg, Wisnioski, \& Wuyts}]{Genzel2020}
Genzel, R., Price, S.~H., {\"U}bler, H., {et~al.} 2020, \apj, 902, 98,
  \dodoi{10.3847/1538-4357/abb0ea}

\bibitem[{Harrison {et~al.}(2018)Harrison, Costa, Tadhunter,
  Fl{\ifmmode\ddot{u}\else\"{u}\fi}tsch, Kakkad, Perna,
  {et~al.}}]{Harrison2018Feb}
Harrison, C.~M., Costa, T., Tadhunter, C.~N., {et~al.} 2018, 2, 198,
  \dodoi{10.1038/s41550-018-0403-6}

\bibitem[{Hopkins \& Elvis(2010)}]{Hopkins2010}
Hopkins, P.~F., \& Elvis, M. 2010, \mnras, 401, 7,
  \dodoi{10.1111/j.1365-2966.2009.15643.x}

\bibitem[{Jeffreson \& Kruijssen(2018)}]{Jeffreson2018}
Jeffreson, S. M.~R., \& Kruijssen, J. M.~D. 2018, \mnras, 476, 3688,
  \dodoi{10.1093/mnras/sty594}

\bibitem[{Kauffmann {et~al.}(2003)Kauffmann, Heckman, White, Charlot, Tremonti,
  Brinchmann, Bruzual, Peng, Seibert, Bernardi, Blanton, Brinkmann, Castander,
  Cs{\'a}bai, Fukugita, Ivezic, Munn, Nichol, Padmanabhan, Thakar, Weinberg, \&
  York}]{Kauffmann2003}
Kauffmann, G., Heckman, T.~M., White, S. D.~M., {et~al.} 2003, \mnras, 341, 33,
  \dodoi{10.1046/j.1365-8711.2003.06291.x}

\bibitem[{Kauffmann {et~al.}(2013)Kauffmann, Pillai, \&
  Goldsmith}]{Kauffmann2013}
Kauffmann, J., Pillai, T., \& Goldsmith, P.~F. 2013, \apj, 779, 185,
  \dodoi{10.1088/0004-637X/779/2/185}

\bibitem[{Kennicutt \& Evans(2012)}]{Kennicutt2012}
Kennicutt, R.~C., \& Evans, N.~J. 2012, \araa, 50, 531,
  \dodoi{10.1146/annurev-astro-081811-125610}

\bibitem[{Kennicutt(1998)}]{Kennicutt1998}
Kennicutt, Jr., R.~C. 1998, \araa, 36, 189,
  \dodoi{10.1146/annurev.astro.36.1.189}

\bibitem[{Kennicutt {et~al.}(2007)Kennicutt, Calzetti, Walter, Helou,
  Hollenbach, Armus, Bendo, Dale, Draine, Engelbracht, Gordon, Prescott, Regan,
  Thornley, Bot, Brinks, de~Blok, de~Mello, Meyer, Moustakas, Murphy, Sheth, \&
  Smith}]{Kennicutt2007}
Kennicutt, Jr., R.~C., Calzetti, D., Walter, F., {et~al.} 2007, \apj, 671, 333,
  \dodoi{10.1086/522300}

\bibitem[{Kewley {et~al.}(2001)Kewley, Dopita, Sutherland, Heisler, \&
  Trevena}]{Kewley2001}
Kewley, L.~J., Dopita, M.~A., Sutherland, R.~S., Heisler, C.~A., \& Trevena, J.
  2001, \apj, 556, 121, \dodoi{10.1086/321545}

\bibitem[{King \& Pounds(2015)}]{King2015}
King, A., \& Pounds, K. 2015, \araa, 53, 115,
  \dodoi{10.1146/annurev-astro-082214-122316}

\bibitem[{Koda \& Sofue(2006)}]{Koda2006}
Koda, J., \& Sofue, Y. 2006, \pasj, 58, 299, \dodoi{10.1093/pasj/58.2.299}

\bibitem[{Kormendy \& Kennicutt(2004)}]{Kormendy2004}
Kormendy, J., \& Kennicutt, Robert~C., J. 2004, \araa, 42, 603,
  \dodoi{10.1146/annurev.astro.42.053102.134024}

\bibitem[{Kreckel {et~al.}(2016)Kreckel, Blanc, Schinnerer, Groves, Adamo,
  Hughes, \& Meidt}]{Kreckel2016}
Kreckel, K., Blanc, G.~A., Schinnerer, E., {et~al.} 2016, \apj, 827, 103,
  \dodoi{10.3847/0004-637X/827/2/103}

\bibitem[{Kristen {et~al.}(1997)Kristen, Jorsater, Lindblad, \&
  Boksenberg}]{Kristen1997}
Kristen, H., Jorsater, S., Lindblad, P.~O., \& Boksenberg, A. 1997, \aap, 328,
  483.
\newblock \url{https://ui.adsabs.harvard.edu/abs/1997A\&A...328..483K}

\bibitem[{Kruijssen {et~al.}(2019)Kruijssen, Schruba, Chevance, Longmore,
  Hygate, Haydon, {et~al.}}]{Kruijssen2019May}
Kruijssen, J. M.~D., Schruba, A., Chevance, M., {et~al.} 2019, Nature, 569,
  519, \dodoi{10.1038/s41586-019-1194-3}

\bibitem[{Krumholz \& McKee(2005)}]{Krumholz2005}
Krumholz, M.~R., \& McKee, C.~F. 2005, \apj, 630, 250, \dodoi{10.1086/431734}

\bibitem[{Krumholz \& Thompson(2007)}]{Krumholz2007}
Krumholz, M.~R., \& Thompson, T.~A. 2007, \apj, 669, 289,
  \dodoi{10.1086/521642}

\bibitem[{Lada {et~al.}(2012)Lada, Forbrich, Lombardi, \& Alves}]{Lada2012}
Lada, C.~J., Forbrich, J., Lombardi, M., \& Alves, J.~F. 2012, \apj, 745, 190,
  \dodoi{10.1088/0004-637X/745/2/190}

\bibitem[{Lada {et~al.}(2010)Lada, Lombardi, \& Alves}]{Lada2010}
Lada, C.~J., Lombardi, M., \& Alves, J.~F. 2010, \apj, 724, 687,
  \dodoi{10.1088/0004-637X/724/1/687}

\bibitem[{Li {et~al.}(2015)Li, Shen, \& Kim}]{Li2015a}
Li, Z., Shen, J., \& Kim, W.-T. 2015, \apj, 806, 150,
  \dodoi{10.1088/0004-637X/806/2/150}

\bibitem[{Lin {et~al.}(2020)Lin, Li, Du, Wang, Xiao, Bureau, Fraser-McKelvie,
  Masters, Lin, Wake, \& Hao}]{Lin2020}
Lin, L., Li, C., Du, C., {et~al.} 2020, \mnras, 499, 1406,
  \dodoi{10.1093/mnras/staa2913}

\bibitem[{Lindblad(1999)}]{Lindblad1999}
Lindblad, P.~O. 1999, \aapr, 9, 221, \dodoi{10.1007/s001590050018}

\bibitem[{Liu {et~al.}(2011)Liu, Koda, Calzetti, Fukuhara, \& Momose}]{Liu2011}
Liu, G., Koda, J., Calzetti, D., Fukuhara, M., \& Momose, R. 2011, \apj, 735,
  63, \dodoi{10.1088/0004-637X/735/1/63}

\bibitem[{Liu {et~al.}(2013)Liu, Calzetti, Hong, Whitmore, Chandar, O'Connell,
  Blair, Cohen, Frogel, \& Kim}]{Liu2013}
Liu, G., Calzetti, D., Hong, S., {et~al.} 2013, \apjl, 778, L41,
  \dodoi{10.1088/2041-8205/778/2/L41}

\bibitem[{Liu {et~al.}(2018)Liu, Wang, Lin, Gao, Liu, Teklu, \& Kong}]{Liu2018}
Liu, Q., Wang, E., Lin, Z., {et~al.} 2018, \apj, 857, 17

\bibitem[{Ly {et~al.}(2014)Ly, Malkan, Nagao, Kashikawa, Shimasaku, \&
  Hayashi}]{Ly2014}
Ly, C., Malkan, M.~A., Nagao, T., {et~al.} 2014, \apj, 780, 122,
  \dodoi{10.1088/0004-637X/780/2/122}

\bibitem[{Maiolino {et~al.}(2017)Maiolino, Russell, Fabian, Carniani,
  Gallagher, Cazzoli, {et~al.}}]{Maiolino2017Mar}
Maiolino, R., Russell, H.~R., Fabian, A.~C., {et~al.} 2017, 544, 202,
  \dodoi{10.1038/nature21677}

\bibitem[{Markwardt(2009)}]{Markwardt2009}
Markwardt, C.~B. 2009, in Astronomical Society of the Pacific Conference
  Series, Vol. 411, Astronomical Data Analysis Software and Systems XVIII, ed.
  D.~A. {Bohlender}, D.~{Durand}, \& P.~{Dowler}, 251.
\newblock \doarXiv{0902.2850}

\bibitem[{McMullin {et~al.}(2007)McMullin, Waters, Schiebel, Young, \&
  Golap}]{McMullin2007}
McMullin, J.~P., Waters, B., Schiebel, D., Young, W., \& Golap, K. 2007,
  Astronomical Society of the Pacific Conference Series, Vol. 376, CASA
  Architecture and Applications, ed. R.~A. {Shaw}, F.~{Hill}, \& D.~J. {Bell},
  127.
\newblock \url{https://ui.adsabs.harvard.edu/abs/2007ASPC..376..127M}

\bibitem[{Momose {et~al.}(2013)Momose, Koda, Kennicutt, Egusa, Calzetti, Liu,
  {et~al.}}]{Momose2013Jul}
Momose, R., Koda, J., Kennicutt, Jr., R.~C., {et~al.} 2013, 772, L13,
  \dodoi{10.1088/2041-8205/772/1/l13}

\bibitem[{Momose {et~al.}(2010)Momose, Okumura, Koda, \& Sawada}]{Momose2010}
Momose, R., Okumura, S.~K., Koda, J., \& Sawada, T. 2010, \apj, 721, 383,
  \dodoi{10.1088/0004-637X/721/1/383}

\bibitem[{Morganti {et~al.}(2015)Morganti, Oosterloo, Oonk, Frieswijk, \&
  Tadhunter}]{Morganti2015}
Morganti, R., Oosterloo, T., Oonk, J. B.~R., Frieswijk, W., \& Tadhunter, C.
  2015, \aap, 580, A1, \dodoi{10.1051/0004-6361/201525860}

\bibitem[{Nguyen-Luong {et~al.}(2016)Nguyen-Luong, Nguyen, Motte, Schneider,
  Fujii, Louvet, Hill, Sanhueza, Chibueze, \& Didelon}]{NguyenLuong2016}
Nguyen-Luong, Q., Nguyen, H. V.~V., Motte, F., {et~al.} 2016, \apj, 833, 23,
  \dodoi{10.3847/0004-637X/833/1/23}

\bibitem[{Onodera {et~al.}(2010)Onodera, Kuno, Tosaki, Kohno, Nakanishi,
  Sawada, Muraoka, Komugi, Miura, Kaneko, Hirota, \& Kawabe}]{Onodera2010}
Onodera, S., Kuno, N., Tosaki, T., {et~al.} 2010, \apjl, 722, L127,
  \dodoi{10.1088/2041-8205/722/2/L127}

\bibitem[{Querejeta {et~al.}(2019)Querejeta, Schinnerer, Schruba, Murphy,
  Meidt, Usero, Leroy, Pety, Bigiel, Chevance, Faesi, Gallagher,
  Garc{\'\i}a-Burillo, Glover, Hygate, Jim{\'e}nez-Donaire, Kruijssen, Momjian,
  Rosolowsky, \& Utomo}]{Querejeta2019a}
Querejeta, M., Schinnerer, E., Schruba, A., {et~al.} 2019, \aap, 625, A19,
  \dodoi{10.1051/0004-6361/201834915}

\bibitem[{Sakamoto {et~al.}(2007)Sakamoto, Ho, Mao, Matsushita, \&
  Peck}]{Sakamoto2007}
Sakamoto, K., Ho, P. T.~P., Mao, R.-Q., Matsushita, S., \& Peck, A.~B. 2007,
  \apj, 654, 782, \dodoi{10.1086/509775}

\bibitem[{Sakamoto {et~al.}(1999)Sakamoto, Okumura, Ishizuki, \&
  Scoville}]{Sakamoto1999}
Sakamoto, K., Okumura, S.~K., Ishizuki, S., \& Scoville, N.~Z. 1999, \apjs,
  124, 403, \dodoi{10.1086/313265}

\bibitem[{Salak {et~al.}(2016)Salak, Nakai, Hatakeyama, \&
  Miyamoto}]{Salak2016}
Salak, D., Nakai, N., Hatakeyama, T., \& Miyamoto, Y. 2016, \apj, 823, 68,
  \dodoi{10.3847/0004-637X/823/1/68}

\bibitem[{Sandqvist {et~al.}(1995)Sandqvist, Joersaeter, \&
  Lindblad}]{Sandqvist1995}
Sandqvist, A., Joersaeter, S., \& Lindblad, P.~O. 1995, \aap, 295, 585

\bibitem[{Sandqvist {et~al.}(1982)Sandqvist, Jorsater, \&
  Lindblad}]{Sandqvist1982}
Sandqvist, A., Jorsater, S., \& Lindblad, P.~O. 1982, \aap, 110, 336

\bibitem[{Schmidt(1959)}]{Schmidt1959}
Schmidt, M. 1959, \apj, 129, 243, \dodoi{10.1086/146614}

\bibitem[{Shin {et~al.}(2019)Shin, Woo, Chung, Baek, Cho, Kang, \&
  Bae}]{Shin2019}
Shin, J., Woo, J.-H., Chung, A., {et~al.} 2019, \apj, 881, 147,
  \dodoi{10.3847/1538-4357/ab2e72}

\bibitem[{Silk(2013)}]{Silk2013}
Silk, J. 2013, \apj, 772, 112, \dodoi{10.1088/0004-637X/772/2/112}

\bibitem[{Sun {et~al.}(2018)Sun, Leroy, Schruba, Rosolowsky, Hughes, Kruijssen,
  Meidt, Schinnerer, Blanc, Bigiel, Bolatto, Chevance, Groves, Herrera, Hygate,
  Pety, Querejeta, Usero, \& Utomo}]{Sun2018a}
Sun, J., Leroy, A.~K., Schruba, A., {et~al.} 2018, \apj, 860, 172,
  \dodoi{10.3847/1538-4357/aac326}

\bibitem[{Usero {et~al.}(2015)Usero, Leroy, Walter, Schruba,
  Garc{\'\i}a-Burillo, Sand~strom, Bigiel, Brinks, Kramer, Rosolowsky,
  Schuster, \& de~Blok}]{Usero2015}
Usero, A., Leroy, A.~K., Walter, F., {et~al.} 2015, \aj, 150, 115,
  \dodoi{10.1088/0004-6256/150/4/115}

\bibitem[{Venturi {et~al.}(2018)Venturi, Nardini, Marconi, Carniani, Mingozzi,
  Cresci, Mannucci, Risaliti, Maiolino, Balmaverde, Bongiorno, Brusa, Capetti,
  Cicone, Ciroi, Feruglio, Fiore, Gallazzi, La~Franca, Mainieri, Matsuoka,
  Nagao, Perna, Piconcelli, Sani, Tozzi, \& Zibetti}]{Venturi2018a}
Venturi, G., Nardini, E., Marconi, A., {et~al.} 2018, \aap, 619, A74,
  \dodoi{10.1051/0004-6361/201833668}

\bibitem[{Wang {et~al.}(2009)Wang, Fabbiano, Elvis, Risaliti, Mazzarella,
  Howell, \& Lord}]{Wang2009}
Wang, J., Fabbiano, G., Elvis, M., {et~al.} 2009, \apj, 694, 718,
  \dodoi{10.1088/0004-637X/694/2/718}

\bibitem[{Wilson {et~al.}(2019)Wilson, Elmegreen, Bemis, \&
  Brunetti}]{Wilson2019}
Wilson, C.~D., Elmegreen, B.~G., Bemis, A., \& Brunetti, N. 2019, \apj, 882, 5,
  \dodoi{10.3847/1538-4357/ab31f3}

\bibitem[{Wu {et~al.}(2005)Wu, Evans, Gao, Solomon, Shirley, \&
  Vanden~Bout}]{Wu2005}
Wu, J., Evans, Neal~J., I.~I., Gao, Y., {et~al.} 2005, \apjl, 635, L173,
  \dodoi{10.1086/499623}

\bibitem[{Xu {et~al.}(2015)Xu, Cao, Lu, Gao, Diaz-Santos, Herrero-Illana,
  Meijerink, Privon, Zhao, Evans, K{\"o}nig, Mazzarella, Aalto, Appleton,
  Armus, Charmandaris, Chu, Haan, Inami, Murphy, Sanders, Schulz, \& van~der
  Werf}]{Xu2015}
Xu, C.~K., Cao, C., Lu, N., {et~al.} 2015, \apj, 799, 11,
  \dodoi{10.1088/0004-637X/799/1/11}

\end{thebibliography}
\bibliographystyle{aasjournal}

\end{document}